%% file: Paper.tex
\documentclass[%
 reprint,
 superscriptaddress,
 nofootinbib,
 amsmath,amssymb,
 aps,
 prd,
 floatfix
]{revtex4-2}

\usepackage{graphicx}% Include figure files
\usepackage{dcolumn}% Align table columns on decimal point
\usepackage{xcolor}
\usepackage{siunitx}
\usepackage{hyperref} % add hypertext capabilities
\input{defs.tex}

\DeclareSIUnit\lightsecond{ls}

\graphicspath{ {./Plots/} {./Figures/} {./figs/} {./} }

\usepackage{soul}

\begin{document}
%% \showthe\columnwidth
\preprint{APS/123-QED}

\title{Generalized parameter-space metrics for continuous gravitational-wave searches}

\author{P. B. Covas}
\email{jb.covas@uib.es}
\affiliation{Max Planck Institute for Gravitational Physics (Albert Einstein Institute) and Leibniz Universit\"at Hannover, D-30167 Hannover, Germany}
\affiliation{Departament de Física, Universitat de les Illes Balears, IAC3 – IEEC, Carretera Valldemossa km 7.5, E-07122 Palma, Spain}
\author{R. Prix}
\affiliation{Max Planck Institute for Gravitational Physics (Albert Einstein Institute) and Leibniz Universit\"at Hannover, D-30167 Hannover, Germany}

%\input{git_tag.tex}

%\date{\commitDATE; \commitIDshort-\commitSTATUS}
%\date{\today}%

\begin{abstract}
  Many searches for continuous gravitational waves face significant computational challenges due to the need to explore large parameter spaces characterized by unknown parameters. Parameter-space metrics are used to predict the relative loss of signal power when the searched parameters differ from the true signal parameters. In this paper we present generalized parameter-space metrics for the $\F$-statistic (a detection statistic used in many searches) that improve upon previous idealized metrics by incorporating realistic effects such as data gaps and varying noise floors. We derive a new marginalized $\F$-statistic metric that is more accurate than the previous averaged $\F$-statistic metric, especially for short coherent segments. We also derive a more accurate semi-coherent metric that properly accounts for the signal-power variability over segments. We provide numerical tests illustrating that the new generalized metrics provide more accurate mismatch predictions than previous expressions. More accurate metrics can result in a reduced number of templates needed for a given search, a feature that could improve the sensitivity of future searches.
\end{abstract}

\maketitle

\section{Introduction}
\label{sec:introduction}

Continuous gravitational waves (CWs) are long-lasting and almost monochromatic gravitational waves (yet to be detected) that can be emitted by different sources, such as asymmetric (around their rotation axis) rotating neutron stars \cite{KeithReview}. Searches for CWs can have different targets, such as known pulsars or unknown neutron stars in our galaxy (commonly called all-sky searches).

In order to discern whether a dataset contains an astrophysical signal or just noise, a detection statistic that is compared to a threshold is commonly computed. One of these detection statistics is the $\F$-statistic \cite{jks98:_data,cutler05:_gen_fstat}, which is obtained after analytical maximization (over the four amplitude parameters that describe a typical CW signal) of the log-likelihood ratio. This detection statistic has been used in many CW searches (see \cite{WETTE2023102880} for a recent review). Due to the prohibitive computational cost of some searches, the data can be divided in many shorter segments and a semi-coherent version of the $\F$-statistic might be used \cite{PhysRevD.61.082001}.

The typical CW signal model includes parameters that for some searches are unknown, such as the rotational frequency or the sky position. These unknown parameters have to be explicitly explored with a template bank or with a stochastic sampling algorithm. The parameter-space metric is a second-order approximation that aims to predict the relative loss of signal power (also called mismatch) incurred when these searched parameters are not equal to the true parameters of a putative astrophysical signal \cite{Metric}.

\begin{figure*}[htbp]
  \begin{center}
    \includegraphics[width=\columnwidth]{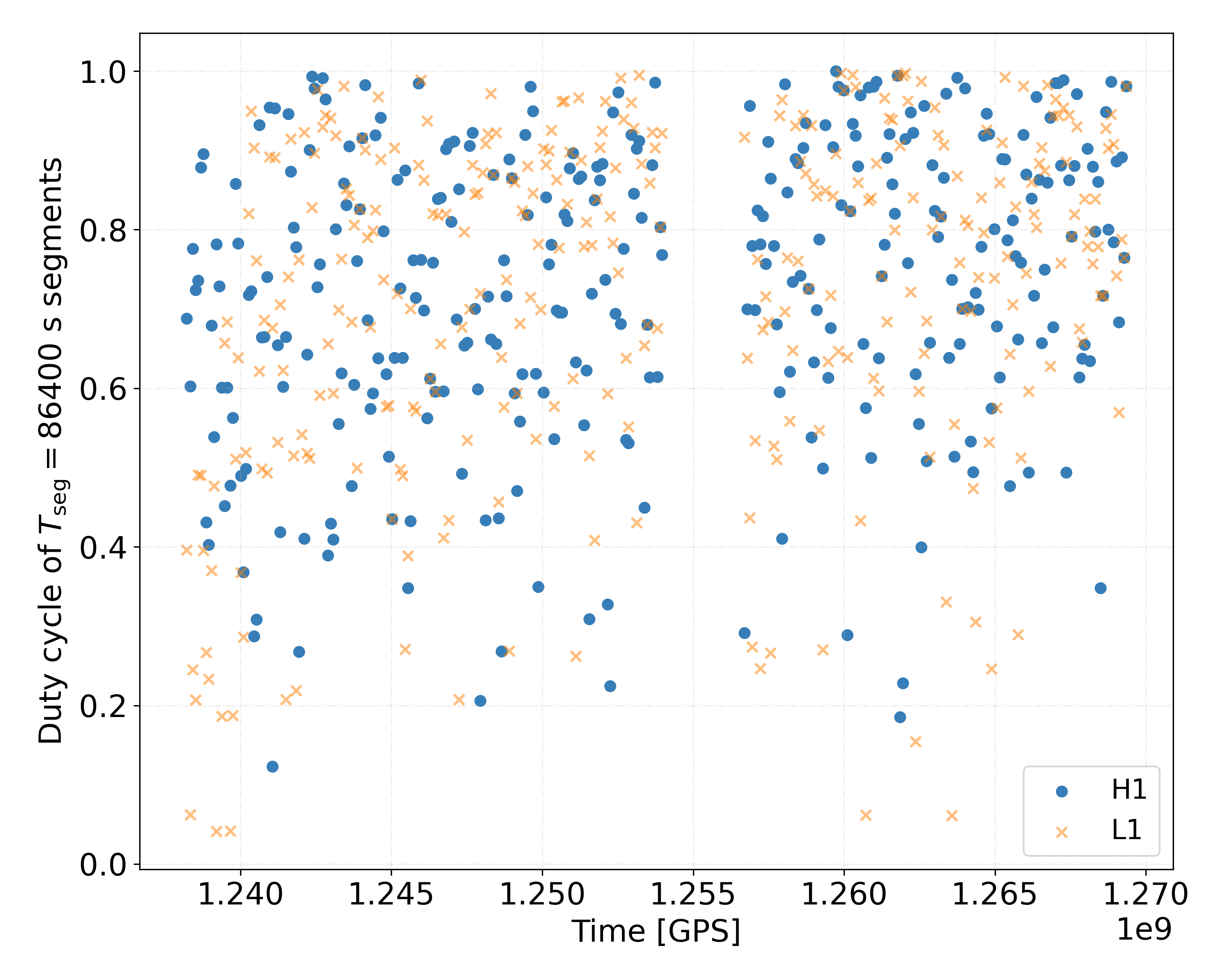}
    \includegraphics[width=\columnwidth]{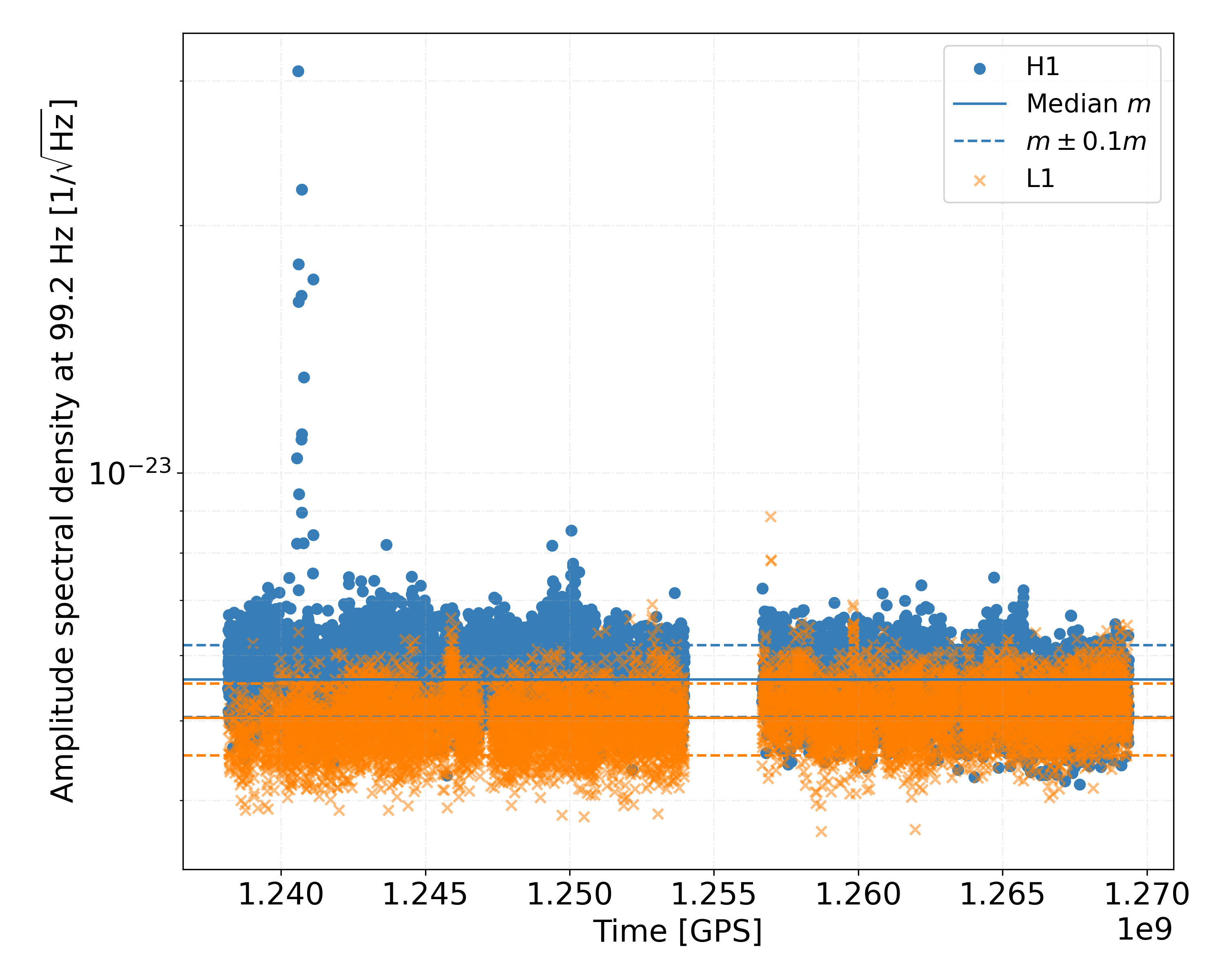}
    \caption{The left plot shows the duty cycle of the O3 observing run for each segment of $\Tseg = \SI{86400}{s}$ as a
      function of the segment mid-time for each detector. The right plot shows the amplitude spectral density (at
      \SI{99.2}{Hz}) of the O3 observing run for each short Fourier transform (SFT) as a function of the SFT mid-time,
      while the horizontal (dashed) lines show the median $m$ ($m \pm 0.1$) for each detector.
      SFTs of \SI{1800}{s} duration are used, which have been cleaned by a time-domain algorithm
      \cite{PhysRevD.105.022005}.
      The different markers show the detectors H1 (blue circles) and L1 (orange crosses).
    }
    \label{fig:motivation}
  \end{center}
\end{figure*}

The full coherent parameter-space metric for the (multi-detector) $\F$-statistic was first derived in \cite{Metric}.
The different metrics derived there are subject to some idealizations, such as
(i) the data is assumed to have no gaps, i.e., a duty cycle of 100$\%$;
(ii) the noise floor of each detector, quantified in terms of the amplitude spectral density (ASD), is assumed to be constant over time.
Furthermore, the semi-coherent metric (cf.~\cite{PhysRevD.61.082001,PhysRevD.82.042002,SemiCoherentMetric})
for the $\F$-statistic has so far been derived under the assumption of (iii) equal signal power over segments.
However, realistic datasets typically have an overall duty cycle less than 100$\%$ (usually around $\sim70\%$),
and varying duty cycles and noise floors over time, as shown in Fig.~\ref{fig:motivation} for the O3 observing
run \cite{GWOSCO3a,GWOSCO3b} of the Advanced LIGO gravitational-wave detectors located in Hanford (H1) and Livingston
(L1) \cite{LIGOScientific:2014pky}.

In this paper we attempt to study the behavior of the $\F$-statistic parameter-space metrics when these assumptions do not hold,
and derive generalized expressions and implementations that take these effects into account.

The parameter-space metric can be used to optimally construct a template grid \cite{Prix_2007,PhysRevD.104.042005}, and since the sensitivity of some CW searches is bounded by a limited computational budget, placing templates in a more accurate way (due to more accurate mismatch predictions) is an important topic of research \cite{PrixShaltev2011..optimalStackSlide}. There are more reasons why obtaining more accurate parameter-space metrics is relevant for CW searches, such as
(i) better analytical estimations of the uncertainty on the unknown parameters (see for example \cite{10.1093/mnras/stab3315}) without having to carry out expensive Bayesian parameter estimation analyses;
(ii) optimal setup algorithms (such as \cite{PrixShaltev2011..optimalStackSlide}) could indicate that discarding some fraction of the less sensitive data can be beneficial, and the resulting dataset would have a lower duty cycle, thus making the usage of more realistic metrics more relevant;
(iii) use jump proposals (within Bayesian stochastic sampling algorithms such as \cite{10.1093/mnras/staa278}) based on the Fisher information matrix, which is proportional to the parameter-space metric;
(iv) compare the resulting posteriors from a Bayesian stochastic sampling analysis with Fisher matrix predictions for validation tests.

The main results of this paper are: (i) we derive an expression for the metric where data gaps and a varying noise floor are taken into account, and we implement this expression numerically; (ii) we present a new expression for a marginalized (over the unknown amplitude parameters) $\F$-statistic metric; (iii) we derive an expression for the semi-coherent metric that properly takes into account the variability of the signal power across different segments by applying weights, and we derive the correct expression for detection statistics that are weighted, such as the one proposed in \cite{WeightedFstat}. We test all of these new expressions and show that they are able to predict the mismatch more accurately than the previous expressions for realistic datasets.

This paper is organized in the following way: in Sec.~\ref{sec:fstat} we give an introduction to the $\F$-statistic and to parameter-space metrics; in Sec.~\ref{sec:new} we derive the expressions for the new generalized parameter-space metrics; in Sec.~\ref{sec:tests} we show the improved accuracy of the new metrics through diverse numerical tests; in Sec.~\ref{sec:end} we summarize the paper, present our conclusions, and advance some ideas for future research.

\section{Background}
\label{sec:fstat}

In this section we introduce the basic framework and notation for the $\F$-statistic and its associated parameter-space metrics.

\subsection{The $\F$-statistic}

In order to detect CWs, we compare two main hypotheses about the data $x(t)$, where $t$ is the time in the detector frame:
\begin{itemize}
 \item noise hypothesis: the data only consists of Gaussian noise, $x(t)=n(t)$.
 \item signal hypothesis: in addition to Gaussian noise, the data consists of a CW signal $s(t;\A\sig,\dop\sig)$
   parametrized by four amplitude parameters $\A\sig$ and a number of phase-evolution parameters $\dop\sig$,
   i.e., $x(t) = n(t) + s(t;\A\sig,\dop\sig)$.
\end{itemize}
The CW signal can be expressed in terms of four basis functions $h_{\mu}$ in the form:
\begin{equation}
\label{eq:signal}
s(t;\A\sig,\dop\sig) = \sum_{\mu=1}^4 \A\sig^{\mu}\,h_{\mu}(\dop\sig),
\end{equation}
where $\A^\mu(h_0,\cosi,\psi,\phi_0)$ are the JKS amplitude parameters \cite{jks98:_data,prix:_cfsv2} depending on the overall signal
amplitude $h_0$, polarization angles $\cosi$ and $\psi$, and the initial phase $\phi_0$. The four basis functions $h_\mu$ are
\begin{align}
  h_1 &= a(t) \cos{\phi(t)}, \quad h_2 = b(t) \cos{\phi(t)}, \\
  h_3 &= a(t) \sin{\phi(t)}, \quad h_4 = b(t) \sin{\phi(t)},
\end{align}
where $a(t)$ and $b(t)$ are the antenna pattern functions (see \cite{jks98:_data,prix:_cfsv2}),
and $\phi(t)$ is the phase of the signal in the detector frame, parameterized by the phase-evolution parameters $\dop$ such as the
frequency $f_0$, spindowns $f_1, f_2, \ldots$, the sky position, and (if applicable) binary orbital parameters.

The log-likelihood ratio between signal and noise hypotheses in this form is found to depend quadratically on the
JKS amplitude parameters $\A^\mu$ of the template and can thus be analytically maximized (reducing the computational cost),
which yields the well-known $\F$-statistic \cite{jks98:_data,cutler05:_gen_fstat,2009CQGra..26t4013P}:
\begin{equation}
    2 \F(x;\dop) = x_{\mu}\,\M^{\mu\nu} x_{\nu}\,,
    \label{eq:2F}
\end{equation}
with implicit summation over repeated $\mu,\nu=1,\ldots,4$.
Here we used the definitions
\begin{equation}
  \label{eq:xmu}
  x_{\mu} \equiv (x|h_\mu), \quad \M_{\mu\nu} \equiv (h_\mu|h_\nu)\,,
\end{equation}
in terms of the multi-detector scalar product \cite{cutler05:_gen_fstat,prix:_cfsv2}, which
for uncorrelated noise between detectors and a narrow-band signal with flat noise floor in the signal band $f$
can be written as
\begin{equation}
  (x|y) \equiv 2 \gamma \left< x y \right>,
  \label{eq:matchedfilter}
\end{equation}
where we have defined the \emph{data factor} $\gamma$ as
\begin{equation}
  \label{eq:18}
  \gamma \equiv \S^{-1}\,\Tdata,
\end{equation}
with the total amount of data $\Tdata \equiv \Nsft\,\Tsft$ in terms of the number $\Nsft$ and duration $T_{\mathrm{SFT}}$ of
short Fourier transforms\footnote{SFTs \cite{Allen_Mendell} are a typical format for the input data of CW searches. We assume stationary noise and constant antenna-pattern coefficients over the duration of each SFT.} (SFTs).
The overall noise floor $\S$ at frequency $f_0$ is defined as
\begin{equation}
  \label{eq:54}
  \S^{-1} \equiv \frac{1}{\Nsft} \sum_{X\alpha} \S_{X\alpha}^{-1},
\end{equation}
in terms of the per-SFT power-spectral densities $\S_{X\alpha}$, where $\alpha$ is an index over SFTs of detector $X$.
We define the time-averaging operation as
\begin{equation}
  \left< x y \right> \equiv \frac{1}{\Nsft} \sum_{X \alpha}^{N_{\mathrm{SFTs}}} \frac{ w_{X \alpha}}{\Tsft} \int\limits_{t_{X \alpha} - \Tsft/2}^{t_{X \alpha} + \Tsft/2} x_X(t) y_X(t) dt,
  \label{eq:1}
\end{equation}
where the timestamps $t_{X\alpha}$ refer to the middle of an SFT.
The per-SFT noise weights $w_{X\alpha}$ in this expression are defined as
\begin{equation}
  \label{eq:5}
  w_{X\alpha} \equiv \frac{\S_{X\alpha}^{-1}}{\S^{-1}}\,,
\end{equation}
with normalization
\begin{align}
  \sum_{X \alpha} w_{X \alpha} &= \Nsft,
\end{align}
as can be seen from Eq.~\eqref{eq:54}.

The $\F$-statistic follows a non-central $\chi^2$-distribution with an expectation value of
\begin{equation}
    \expect{2 \F(x;\dop)} = 4 + \rho^2(\A\sig,\dop\sig;\dop),
    \label{eq:2Fexp}
\end{equation}
with the non-centrality parameter defining the \emph{signal power} $\rho^{2}(\A\sig,\dop\sig;\dop)$.
In the perfect-match case, where the phase-evolution
parameters of the template $\dop$ match those of the signal $\dop\sig$,
the signal power can be expressed \cite{PhysRevD.98.084058} as:
\begin{align}
  \rho^2_0 &\equiv \rho^2(\A\sig,\dop\sig;\dop\sig) = (s|s) = h_0^2 \gamma\,g^2,
  \label{eq:rho2}
\end{align}
where we used the definition of the geometric response function $g$ of \cite{prix2024analyticweaksignalapproximationbayes}, namely
\begin{align}
  g^2(\vn,\cosi,\psi) &\equiv \alpha_1\,A + \alpha_2\,B + 2 \alpha_3\,C ,
  \label{eq:g}
\end{align}
where $\vn$ is the sky position, the amplitude angle factors $\alpha_k(\cosi,\psi)$ are defined as:
\begin{equation}
  \label{eq:51}
  \begin{aligned}
    \alpha_1 &\equiv \frac{1}{4}\left(1+\cosisq\right)^{2} \cos ^{2}2\psi+\cosisq\,\sin ^{2}2 \psi \,, \\
    \alpha_2 &\equiv \frac{1}{4}\left(1+\cosisq\right)^{2} \sin ^{2}2\psi+\cosisq\,\cos ^{2}2 \psi \,, \\
    \alpha_3 &\equiv \frac{1}{4}\left(1-\cosisq\right)^{2} \sin 2 \psi\, \cos 2 \psi,
  \end{aligned}
\end{equation}
and $\gamma\,A \equiv \M_{11}=\M_{33}$, $\gamma\,B \equiv \M_{22}=\M_{44}$, and $\gamma\,C\equiv \M_{12}=\M_{34}$
are the (non-negligible) components of the (symmetric) antenna-pattern matrix $\M_{\mu\nu}$ defined in Eq.~\eqref{eq:xmu}, with
\begin{align}
  A = \left< a^2 \right>,\;
  B = \left< b^2 \right>,\;
  C = \left< a b \right>,
\end{align}
and the sub-determinant $D \equiv AB - C^2$.

Due to the high computational cost of wide-parameter-space CW searches, these typically use semi-coherent methods, where the data
will be divided into $N\seg$ segments of duration $T\seg$ each, and the four signal amplitude parameters $\A^\mu$ are required to be constant
only within each segment, resulting in the semi-coherent $\sco{\F}$-statistic \cite{PhysRevD.61.082001,prix_search_2011}
as a sum of coherent $\F$-statistics over segments.
This was generalized in \cite{WeightedFstat} to the \emph{weighted} semi-coherent $\sco{\F}_\weighted$-statistic, defined as
\begin{equation}
  \label{eq:10}
  2\sco{\F}_{\weighted}(x;\dop) \equiv \sum_{\ell=1}^{N\seg} \weight_\ell\, 2\F_\ell(x;\dop)\,,
\end{equation}
where $\F_\ell$ is the coherent $\F$-statistic in segment $\ell$ and $\weight_\ell$ is the per-segment weight,
with standard normalization (to have unit mean):
\begin{equation}
  \label{eq:3}
  \sum_{\ell=1}^{N\seg} \weight_\ell = \Nseg.
\end{equation}
Dropping the per-segment weights recovers the classic semi-coherent $\F$-statistic $2\sco{\F}(x;\dop)$.
As discussed in \cite{WeightedFstat}, a good choice of weights is empirically  found as
\begin{equation}
  \label{eq:2}
  \weight_\ell = k\,\gamma_\ell \left( A_\ell + B_\ell \right),
\end{equation}
where the normalization $k$ is fixed by Eq.~\eqref{eq:3}.
This (weighted) semi-coherent $\F$-statistic follows a (\emph{generalized}) non-central $\chi^2$-distribution with an expectation value of (assuming that all segments have four degrees of freedom)
\begin{align}
  \expect{2\sco{\F}_\weighted(x;\dop)} &= \sum_{\ell=1}^{N\seg} \weight_\ell ( 4 + \rho^2_\ell)\nonumber\\
%                                   & = 4 N\seg + \sum_{\ell=1}^{N\seg} \weight_\ell \rho^2_\ell \nonumber \\
                                   &= 4 N\seg + \sco{\rho}_\weighted^2,  \label{eq:2Fswexp}
\end{align}
where $\rho^2_\ell$ is the signal power in segment $\ell$ and we have defined the (weighted) semi-coherent signal power
$\sco{\rho}_\weighted^2$ as
\begin{align}
  \sco{\rho}_\weighted^2 \equiv \sum_{\ell=1}^{N\seg} \weight_\ell\rho^2_\ell ,
  \label{eq:rho2w}
\end{align}
which reduces to the unweighted semi-coherent signal power $\sco{\rho}^2$ when dropping the weights $\weight_\ell$.

\subsection{Parameter-space metrics}
\label{sec:metric}

Due to the high computational cost of wide parameter-space CW searches, a grid of templates with finite spacing is used to cover the selected parameter-space region. For this reason, the values of the searched parameters will not be equal to the parameters of a possible astrophysical signal.
The (coherent) mismatch $\mu$ describes the relative loss of signal power $\rho_{0}^2\equiv\rho^2(\A\sig,\dop\sig;\dop\sig)$ due to computing the coherent detection statistic at offset phase-evolution parameters $\dop = \dop\sig + \Delta\dop$ with offset signal power $\rho_{\ddop}^2\equiv\rho^2(\A\sig,\dop\sig;\dop)$, i.e.\
\begin{align}
    \mu \equiv 1 - \frac{\rho_{\ddop}^2}{\rho^2_{0}},
    \label{eq:mismatchsegment}
\end{align}
which ranges from $0$ (fully recovered signal power) to $1$ (no recovered signal power).

The mismatch $\mu$ can be approximated by a Taylor expansion of the signal power around the signal parameters (where the
mismatch attains a minimum of 0), keeping terms only up to second order \cite{Metric}: 
\begin{equation}
  m \equiv g_{ij}\,\Delta\dop^i\Delta\dop^j\,,\quad
  \mu \approx m + \Ord{\Delta\dop^3},
  \label{eq:mismatch-taylor}
\end{equation}
where $g_{ij}$ is the parameter-space \emph{metric} and $i$ and $j$ are indices over the phase-evolution parameters.
The true mismatch $\mu$ of Eq.~\eqref{eq:mismatchsegment} is bound within $[0, 1]$, while the metric-mismatch
prediction $m$ is quadratic in offsets $\Delta\dop$ and can therefore be larger than one.

It has been shown that the metric mismatch $m$ is typically a good approximation of the true mismatch $\mu$
of Eq.~\eqref{eq:mismatchsegment} up to mismatches of $\mu \lesssim 0.1$ \cite{Metric,wette_flat_2013},
above which it starts to overestimate the true mismatch.
The metric is useful to build template banks of given maximal mismatch and one can estimate the resulting number of
templates $\mathcal{N}$ needed to cover a parameter-space region $\mathcal{R}$ by an expression of the form (neglecting boundary effects)
\cite{PhysRevD.97.103020}:
\begin{align}
  \mathcal{N} \propto \int_{\mathcal{R}}  \sqrt{ \mathrm{det} (g_{ij}) } d \lambda.
  \label{eq:nstar}
\end{align}

The most general form of the metric, as defined by Taylor-expanding Eq.~\eqref{eq:mismatchsegment}, results in the
$\F$-statistic metric $g_{ij}^{\F}(\cosi,\psi;\dop)$, which
depends on the subset $\{\cosi,\psi\}$ of the four signal amplitude parameters $\A\sig$.
An explicit expression for this $\F$-statistic metric can be obtained \cite{Metric} as:
\begin{equation}
  g_{ij}^{\F}(\cosi,\psi;\dop) = \frac{ \alpha_1\, m_{1;ij} + \alpha_2\, m_{2;ij} + 2 \alpha_3\, m_{3;ij} }{ g^2 } ,
  \label{eq:metric-fstat}
\end{equation}
in terms of the amplitude angle functions $\alpha_i(\cosi,\psi)$ defined in Eq.~\eqref{eq:51},
and the matrices
\begin{widetext}
\begin{align}
  \label{eq:m1}
  m_{1;ij} &\equiv \left< a^2 \partial_i \phi \partial_j \phi \right> - \frac{A \left< a b \partial_i \phi \right> \left< a b \partial_j \phi \right> + B \left< a^2 \partial_i \phi \right> \left< a^2 \partial_j \phi \right> - 2 C \left< a^2 \partial_i \phi \right> \left< a b \partial_j \phi \right>}{D} \,, \nonumber\\
  m_{2;ij} &\equiv \left< b^2 \partial_i \phi \partial_j \phi \right> - \frac{A \left< b^2 \partial_i \phi \right> \left< b^2 \partial_j \phi \right> + B \left< a b \partial_i \phi \right> \left< a b \partial_j \phi \right> - 2 C \left< ab \partial_i \phi \right> \left< b^2 \partial_j \phi \right>}{D} \,, \\
  m_{3;ij} &\equiv \left< a b \partial_i \phi \partial_j \phi \right> - \frac{A \left< a b \partial_i \phi \right> \left< b^2 \partial_j \phi \right> + B \left< ab \partial_i \phi \right> \left< a^2 \partial_j \phi \right>  - C \left[ \left< b^2 \partial_i \phi \right> \left< a^2 \partial_j \phi \right> + \left< a b \partial_i \phi \right> \left< a b \partial_j \phi \right> \right]}{D} ,\nonumber
  \label{eq:m3}
\end{align}
\end{widetext}
where $\partial_i \phi \equiv \frac{\partial \phi}{\partial \lambda^i }$ are the partial phase derivatives for the phase-evolution parameters $\dop^i$.

The signal amplitude parameters $\cosi$ and $\psi$ are generally unknown, therefore the practical usefulness of this metric
is somewhat limited.
One level of simplification consists in an amplitude-``average'' form of the $\F$-statistic metric, $g^{\meanF}_{ij}$, which was obtained
\cite{Metric} in the form
\begin{equation}
  g_{ij}^{\meanF}(\dop) = \frac{B\, m_{1;ij} + A\, m_{2;ij} - 2 C\, m_{3;ij} }{2 D} .
  \label{eq:metric-fstat-average}
\end{equation}
However, the most commonly-used approximation is the \emph{phase metric} $g_{ij}^{\phi}$
(initially proposed in \cite{PhysRevD.57.2101}), which is obtained by neglecting the signal amplitude modulation,
resulting in the simpler expression
\begin{align}
  g_{ij}^{\phi}(\dop) = \avg{\partial_i \phi\, \partial_j \phi} - \avg{\partial_i \phi} \avg{ \partial_j \phi}.
  \label{eq:metric-phase}
\end{align}
The semi-coherent metric is typically obtained by averaging the coherent metrics over segments, namely
\begin{align}
  \sco{g}_{ij} = \frac{1}{\Nseg}\, \sum_\ell g_{ij,\ell},
  \label{eq:semicohphasemetric}
\end{align}
which can be derived under the assumption of \emph{constant per-segment signal power}
(see for example \cite{SemiCoherentMetric} or \cite{PhysRevD.82.042002}).

\section{Generalized parameter-space metrics}
\label{sec:new}

In this section we derive more general expressions for the various parameter-space metrics and discuss improvements in
their implementation, in order to take into account data gaps and non-stationarities.
This results in more accurate and robust parameter-space metrics, which will be tested numerically in Sec.~\ref{sec:tests}.

\subsection{Data gaps and varying noise floor}

The metric expressions in Eqs.~\eqref{eq:metric-fstat}, \eqref{eq:metric-fstat-average}, and \eqref{eq:metric-phase} are
typically computed numerically.
However, current metric implementations in the $\F$-statistic context, including the main implementation in
\textit{UniversalDopplerMetric} of \textit{lalsuite} \cite{lalsuite},
tend to implement the average Eq.~\eqref{eq:1} as a simple (unweighted) time-average over the full observation span $T$ of a segment, i.e.,
$\avg{Q}\approx 1/T\int_0^TQ(t)dt$,
without taking into account data gaps or varying noise floors\footnote{However, per-SFT summing and noise-weighting
  for data gaps and non-constant noise floor \emph{has} been used previously for the
  cross-correlation statistic, see Sec.~IV of \cite{PhysRevD.91.102005}.}.
There is some support for differing noise-levels between detectors (see Eq.~(59) in \cite{Metric}), but this does not include
noise-floor variations over time and is only implemented in the $\F$-statistic metrics $g^{\F}$ and $g^{\meanF}$,
not the phase metric $g^{\phi}$.

In order to take into account data gaps and varying noise floors in the metric, we only need to use the full expression
Eq.~\eqref{eq:1} for the weighted per-SFT average, which already incorporates these effects.
We can see that SFTs with a higher noise floor $\S_{X\alpha}$ will contribute less to the metric integral.
Therefore adding noisier data (e.g., from a less sensitive previous observing run) will not improve the
parameter-space resolution as much as the idealized metrics would predict (e.g., a linear scaling with the observation time for
frequency resolution, for example, see Eq.~(43) of \cite{PhysRevD.93.064011}).
The effects of this more general implementation will be illustrated in Sec.~\ref{sec:tests}.

\subsection{New \emph{marginalized} $\F$-statistic metric}
\label{sec:new-marginalized-f-metric}
As discussed in Sec.~\ref{sec:metric}, the ``average'' $\F$-statistic metric $g^{\meanF}$ in \cite{Metric} is obtained as
the midpoint between mismatch extrema of $g^{\F}$ over $\cosi,\psi$.
A more natural approach consists of marginalizing the $\F$-statistic metric $g^{\F}(\cosi,\psi;\dop)$ over the unknown
amplitude parameters $\{\cosi,\psi\}$ using their physical ignorance priors $P(\cosi,\psi)=1/\pi$,
corresponding to isotropic axis orientation of the CW source (e.g., see \cite{2009CQGra..26t4013P}).

We define the \emph{marginalized} $\F$-statistic metric $g^{\margF}$ as
\begin{align}
  \label{eq:4}
  g^{\margF}_{ij}(\dop) &\equiv \avg{g^{\F}_{ij}}_{\cosi,\psi} \nonumber\\
                     &= \frac{1}{\pi}\int_{-1}^{1}d\cosi\int_{-\pi/4}^{\pi/4}d\psi\, g^{\F}(\cosi,\psi;\dop),
\end{align}
which can be approximated (see Sec.~\ref{appendix:averaged}) as
\begin{equation}
  \label{eq:metric-marginalized}
  g_{ij}^{\margF} \approx \frac{m_{1;ij} + m_{2;ij}}{A+B}.
\end{equation}
This expression can be substantially more accurate than the ``average'' $\F$-statistic metric $g^{\meanF}$, especially
for short coherent segments, as shown in Sec.~\ref{sec:monte-carlo-simul}, particularly Fig.~\ref{fig:semicoherent2}.

The marginalized metric $g^{\margF}$ does not involve the inverse (sub)-determinant $D^{-1}$, contrary to the ``average''
$g^{\meanF}$ of Eq.~\eqref{eq:metric-fstat-average}, which can lead to numerical problems for short segments when
$D\rightarrow 0$ (see \cite{PhysRevD.105.124007}).

\subsection{Generalized semi-coherent metric}
\label{sec:semi-coherent}

The usual expression Eq.~\eqref{eq:semicohphasemetric} for the semi-coherent metric, which was originally derived in
\cite{PhysRevD.61.082001} and is commonly found in the literature (e.g., see \cite{PhysRevD.82.042002,SemiCoherentMetric}),
is derived under the assumption of constant signal power $\rho^2_\ell$ over all segments.
However, this is not generally a good approximation due to varying
(i) antenna-patterns $A,B,C$,
(ii) data amounts $\Tdata$ and
(iii) noise-floors $\S$ over segments,
as seen from Eq.~\eqref{eq:rho2} and Fig.~\ref{fig:motivation}.

A more accurate semi-coherent metric can be derived by relaxing this assumption.
The following derivation is written in terms of the general \emph{weighted} semi-coherent $\sco{\F}_\weighted$-statistic
of Eq.~\eqref{eq:10}, as the unweighted special case can be recovered by simply dropping the per-segment weights $\weight_\ell$.

Starting from the general mismatch definition given by Eq.~\eqref{eq:mismatchsegment}, the semi-coherent mismatch
$\sco{\mu}_\weighted$ can be defined as%
\footnote{Note that \emph{interpolating} semi-coherent statistics require adjustments to the total mismatch, see
  \cite{PrixShaltev2011..optimalStackSlide,SemiCoherentMetric}, but those can be equally used with the expressions
  derived here.}
\begin{equation}
  \label{eq:7}
  \sco{\mu}_\weighted \equiv 1 - \frac{\sco{\rho}_{\weighted\ddop}^2}{\sco{\rho}_{\weighted 0}^2},
\end{equation}
in terms of the offset and perfect-match semi-coherent signal power of Eq.~\eqref{eq:rho2w} using either
offset $\rho_{\ddop;\ell}^2$ or perfect-match per-segment signal power $\rho_{0;\ell}^2$, respectively.
Using the per-segment mismatch $\mu_\ell$ of Eq.~\eqref{eq:mismatchsegment}, we can express the offset per-segment
signal power as
\begin{equation}
  \label{eq:6}
  \rho^2_{\ddop;\ell} = (1-\mu_\ell)\,\rho^2_{0;\ell},
\end{equation}
therefore
\begin{equation}
  \label{eq:8}
  \sco{\rho}_{\weighted \ddop}^2 = \sum_\ell \weight_\ell\rho_{\ddop;\ell}^2 = \sco{\rho}_{\weighted 0}^2 - \sum_\ell \weight_\ell\rho_{0;\ell}^2\,\mu_\ell\,,
\end{equation}
and substituting into Eq.~\eqref{eq:7} we obtain
\begin{equation}
  \label{eq:mu_w}
  \sco{\mu}_\weighted = \frac{1}{\Nseg}\sum_{\ell=1}^{\Nseg} W_\ell\,\mu_\ell\,,
\end{equation}
where we defined metric segment weights $W_\ell$ as
\begin{equation}
  \label{eq:11}
  W_\ell \equiv \frac{\Nseg \,\weight_\ell \,\rho_{0;\ell}^2}{\sco{\rho}_{\weighted 0}^2},
\end{equation}
with the usual weight normalization of $\sum_\ell W_\ell = \Nseg$.
As mentioned before, the ``classic'' semi-coherent metric expression given by Eq.~\eqref{eq:semicohphasemetric} can be recovered
under the assumption of constant (weighted) signal power over segments, i.e., $W_\ell=1$.

Applying the metric expansion of Eq.~\eqref{eq:mismatch-taylor} for the $\F$-statistic metric
$g^{\F}_{ij,\ell}$, we find
\begin{align}
  \sco{\mu}_\weighted &\approx \sco{g}^{\F}_{\weighted;ij}\,\Delta\dop^i\Delta\dop^j + \Ord{\Delta\dop^3},\quad\text{with}   \label{eq:13a} \\
  \sco{g}^{\F}_{\weighted;ij} &\equiv \frac{1}{\Nseg}\sum_\ell W_\ell\, g^{\F}_{ij,\ell}\,. \label{eq:13b}
\end{align}
As noted before in Sec.~\ref{sec:metric}, however, this expression is not typically usable in practice, because both the metric
weights $W_\ell$ and the $\F$-statistic metric $g^{\F}$ depend on the unknown signal amplitude parameters
$\{\cosi,\psi\}$.
The most natural approach again is to marginalize over them using physical priors, as discussed as in
Sec.~\ref{sec:new-marginalized-f-metric}, and using the same marginalization approximation as before
(see Sec.~\ref{appendix:semicohw} for details), we obtain the general result
\begin{equation}
  \label{eq:14}
  \sco{g}_{\weighted;ij}(\dop) \approx \frac{1}{\Nseg} \sum_{\ell=1}^{\Nseg}\mweight_\ell \,g_{ij,\ell}(\dop)\,,
\end{equation}
in terms of the marginalized \emph{metric segment weights}
\begin{equation}
  \label{eq:15}
  \mweight_\ell \propto \weight_\ell\,\gamma_\ell\left(A_\ell + B_\ell\right)\,,
\end{equation}
with standard weight normalization $\sum_\ell \mweight_\ell = \Nseg$.

There are two important cases to distinguish, namely whether the underlying semi-coherent statistic is weighted
by $\weight_\ell$ or not (i.e., dropping $\weight_\ell$ in the above expressions).
The explicit metric weights for two cases are therefore
\begin{equation}
  \label{eq:16}
  \mweight_\ell = \begin{cases}
    k' \weight_\ell^2 & \text{for weighted } \sco{\F}_{\weighted}\,,\\
    \weight_\ell & \text{for unweighted } \sco{\F}\,,
  \end{cases}
\end{equation}
in terms of the segment weights $\weight_\ell$ given in
Eq.~\eqref{eq:2}.

For the unweighted $\sco{\F}$-statistic the metric weights are therefore equal to the statistic weights $\weight_\ell$
derived in \cite{WeightedFstat}, while the metric weights for the weighted $\sco{\F}_{\weighted}$-statistic are
proportional to the \emph{squared} statistic weights.

A potential practical difficulty introduced by these metric weights is an additional sky-position dependency (via the
antenna-pattern coefficients $A,B$), even if the per-segment metric itself was constant over the sky.
This can complicate template-bank placement, but one would likely still reap benefits from using sky-marginalized weights instead,
namely $\avg{\mweight_\ell}_{\mathrm{sky}}\propto \{\gamma_\ell^2, \gamma_\ell\}$ (for weighted and unweighted statistics,
respectively), therefore still accounting for the variations in data-quality and quantity over segments illustrated in
Fig.~\ref{fig:motivation}.

\section{Numerical tests}
\label{sec:tests}
\newcommand{\sskip}{0.2em}
\newcommand{\bskip}{0.4em}

In this section we compare the accuracy of the (new) generalized parameter-space metrics derived in
Sec.~\ref{sec:new} to the (previous) idealized metrics, which ignore data gaps, non-constant noise floors, and
signal-power variability between segments.
All calculations and simulations in this section use the C99 \textsc{LALSuite} library via the
\textsc{SWIGLAL} python interface \cite{swig}.

\subsection{Visual illustration}
\label{sec:visual-illustration}
\begin{figure*}[htbp]
    \parbox{0.32\textwidth}{
      (i) coherent $\F$-statistic:\\
      \includegraphics[width=0.33\textwidth]{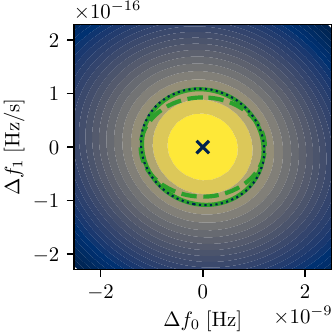}
    }
    \parbox{0.32\textwidth}{
      \hspace*{-2em}(ii) semi-coherent $\sco{\F}$:\\
      \includegraphics[width=0.33\textwidth]{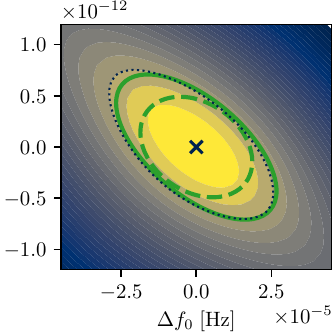}
    }
    \parbox{0.32\textwidth}{
      (iii) semi-coherent {weighted} $\sco{\F}_\weighted$:\\
      \includegraphics[width=0.33\textwidth]{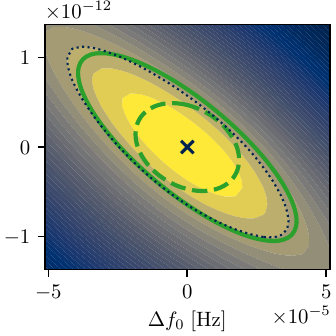}
    }
    \caption{Contour levels of \emph{measured} mismatch $\mu$ in frequency $f_0$ and spindown $f_1$ around an injected
      signal marked by ``$\times$'' (with parameters $\cosi = 0.43$, $\psi = \SI{0.01}{\radian}$, $\alpha = \SI{0.14}{\radian}$,
      $\delta = \SI{-0.27}{\radian}$, $f_0=\SI{100}{\hertz}$, $f_1=\SI{6.4e-12}{\hertz/\second}$ at GPS time
      $\SI{1218673431.5}{\second}$) in O2+O3 data.
      Overlaid: (dotted dark blue) contour-line for measured-mismatch level $\mu=0.05$ and the corresponding mismatch
      ellipses at $m=0.05$ for the $\F$-statistic metric $g^{\F}$ using the previous idealized (dashed green)
        and the new generalized (solid green) metric, for
        (i) coherent $\F$-statistic,
        (ii) semi-coherent unweighted $\sco{\F}$-statistic, and
        (iii) semi-coherent weighted $\sco{\F}_\weighted$-statistic.
        Both semi-coherent cases use short data segments of duration $\Tseg = \SI{0.1}{\day}$.
      }
    \label{fig:heatmaps}
\end{figure*}
As a visual illustration of the improvements, Fig.~\ref{fig:heatmaps} shows contour-levels of measured mismatch $\mu$
around an injected signal, overlaid with the metric ellipse for the (previous) idealized and the (new) generalized
$\F$-statistic metric $g^\F$ of Eq.~\eqref{eq:metric-fstat}, computed for the exact injected signal parameters.
The figure shows three examples, a fully coherent and two semi-coherent short-segment $\Tseg=\SI{0.1}{\day}$ statistics,
using either unweighted or weighted $\F$-statistics.
In all three cases we observe improved agreement between the mismatch-ellipses of the new metrics with the
corresponding measured-mismatch contour level.

\subsection{Monte-Carlo tests of metric-prediction error}
\label{sec:monte-carlo-simul}

In order to quantify the metric-prediction error, we compute the (symmetric) relative error $\varepsilon$ between
metric mismatch prediction $m$ of Eq.~\eqref{eq:mismatch-taylor} and the true mismatch $\mu$ of
Eq.~\eqref{eq:mismatchsegment} (following \cite{Metric,SemiCoherentMetric,2015arXiv150200914L}) as
\begin{equation}
  \label{eq:9}
  \varepsilon \equiv 2 \, \frac{ \mu - m }{ \mu + m } \in (-2, 2).
\end{equation}
Note that the metric tends to increasingly over-estimate the true mismatch $\mu$ at increasing mismatches
$\mu \gtrsim 0.1$ (e.g., see \cite{Metric,wette_flat_2013}), therefore these errors will typically have a bias towards
negative values.

We compare the previous, idealized metric predictions to the new, generalized method of Sec.~\ref{sec:semi-coherent} for
the phase metric $g^\phi$, the averaged $\F$-statistic metric $g^{\meanF}$, and the full $\F$-statistic metric $g^{\F}$.
Additionally we show the prediction errors of the new marginalized $\F$-statistic metric $g^{\margF}$ that has no
previous equivalent.

For these tests we generate \num{100} simulated CW signals with random amplitude parameters drawn from their
physical priors (see Sec.~\ref{sec:new-marginalized-f-metric}), isotropically distributed sky positions (using
equatorial coordinates $\alpha$ and $\delta$), frequency and spindowns drawn uniformly from
$f_0\in [99.9, 100.1]~\si{Hz}$, $f_1 \in [\num{-e-10}, \num{e-10}]~\si{Hz/s}$, $f_2 = [\num{-e-17}, \num{e-17}]~\si{Hz/s^2}$,
and binary parameters: projected semi-major axis $\asini \in [10, 40]~\si{\lightsecond}$, period $\Porb \in [15, 45]~\si{\day}$,
eccentricity $\ecc \in [0, 0.1]$, and argument of periapsis $\argp \in [0, 2\pi)$.
We fix the time of ascending node $\tasc$ and the reference time for frequency and spindowns to the mid-time of the dataset.

For each signal we generate \num{100} parameter-space offsets $\Delta\lambda$, drawn uniformly within the $\F$-statistic
metric ellipsoid $g^{\F}_{ij}(\lambda\sig)\,d\lambda^i d\lambda^j \le 0.2$ (cf.\ Sec.~IV C of
\cite{PhysRevD.110.024053}), computed at each signal location $\lambda\sig$ (resulting in a total of \num{10000} offsets
$\Delta\lambda$).

Given an offset, we can measure the mismatch $\mu(\Delta\lambda)$ of Eq.~\eqref{eq:mismatchsegment} between the
statistic at the signal parameters $\lambda\sig$ and the offset point $\lambda\sig+\Delta\lambda$, and compare this to
the metric prediction $m(\Delta\lambda)=g_{ij}(\lambda\sig)\,\Delta\lambda^i\Delta\lambda^j$ via the relative error Eq.~\eqref{eq:9}.

We consider five different search statistics, namely fully coherent $\F$, long-segment ($\Tseg=\SI{10}{\day}$) semi-coherent
and short-segment ($\Tseg=\SI{0.1}{\day}$) semi-coherent, both for the unweighted $\sco{\F}$ and weighted
$\sco{\F}_\weighted$ statistics.
For each search statistic we consider four different search targets:
\begin{itemize}
\item[(a)] \emph{directed isolated} with offsets in $\{ f_0, f_1, f_2 \}$,
\item[(b)] \emph{directed binary} with offsets in $\{ f_0, \asini, \Porb, \tasc \}$,
\item[(c)] \emph{all-sky isolated} with offsets in $\{ f_0, f_1, \alpha, \delta \}$, and
\item[(d)] \emph{all-sky binary} with offsets in $\{ f_0, \alpha, \delta, \asini, \Porb, \tasc \}$.
\end{itemize}
We use public data from the O2 and O3 observing runs \citep{GWOSCO2,GWOSCO3a,GWOSCO3b} of the Advanced LIGO detectors to
construct realistic datasets in terms of their distribution of SFT timestamps and noise ASD values.
Each of the simulated CW signals is generated with duty cycles and ASD values corresponding to those O2/O3 datasets, but
since we generate signals without noise (as required for metric tests), the fluctuating ASD values are used as
\emph{assumed} noise-floors when computing the $\F$-statistics (via Eq. \eqref{eq:2F}).

\subsubsection{Coherent search}

\begin{figure*}[htbp]
  \textbf{Coherent ${\F}$-statistic:}\\[\bskip]
  \includegraphics[width=\columnwidth]{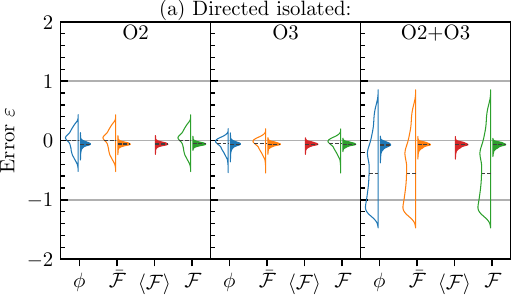}
  \includegraphics[width=\columnwidth]{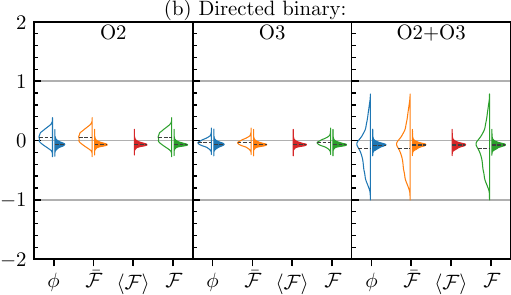}\\[\sskip]
  \includegraphics[width=\columnwidth]{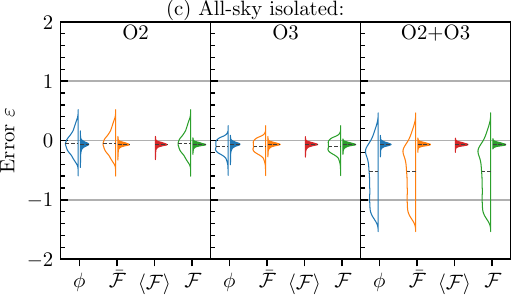}
  \includegraphics[width=\columnwidth]{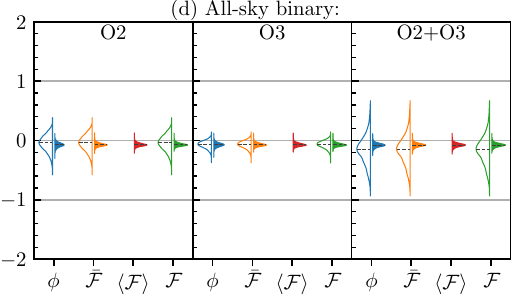}
  \caption{
    Distributions of relative errors $\varepsilon$ of Eq.~\eqref{eq:9} for metric mismatch predictions for different
    coherent $\F$-statistic searches:
    (a) directed isolated, (b) directed binary, (c) all-sky isolated, and (d) all-sky binary,
    using different data-sets: O2, O3, or O2+O3-combined, and metrics:
    phase metric $g^\phi$, averaged $\F$-statistic metric $g^{\meanF}$, marginalized $\F$-statistic metric $g^{\margF}$,
    and the full $\F$-statistic metric $g^{\F}$.
    Each half-violin shows the error distribution over $\Ord{10^4}$ injections for one metric, where the left half
    (empty) uses the previous \emph{idealized} metric while the right half (filled) uses the new \emph{generalized}
    metric implementation discussed in Sec.~\ref{sec:new}. The dashed vertical lines indicate the distribution median.
  }
  \label{fig:coherent}
\end{figure*}
For the fully coherent $\F$-statistic search the resulting distributions of the metric prediction error $\varepsilon$
compared to the measured mismatch are shown in Fig. \ref{fig:coherent}.
We observe a general improvement of the metric prediction errors in the new generalized metric implementation compared
to the previous idealized one, as the distribution mass is closer to zero (with a slight negative bias due to the known
metric-mismatch overestimate mentioned earlier).
Especially the big data gap between O2 and O3 causes large errors in the idealized metric computation for O2+O3, where
this gap cannot be properly taken into account.
We also note that all four different (generalized) metrics (i.e.,
$g^{\phi}$, $g^{\meanF}$, $g^{\margF}$, and $g^{\F}$) show very similar errors in all cases, which can be understood from
the very long coherence time, where the $\F$-statistic metric is expected to converge to the phase-metric as
previously discussed in \cite{Metric,SemiCoherentMetric}.

\subsubsection{Long-segment $\Tseg=\SI{10}{\day}$ semi-coherent search}

\begin{figure*}[htbp]
  % --- First 2x2 block: Unweighted ---
  \textbf{I.\ Unweighted $\sco{\F}$-statistic ($\Tseg=\SI{10}{\day}$):}\\[\bskip]
  \includegraphics[width=\columnwidth]{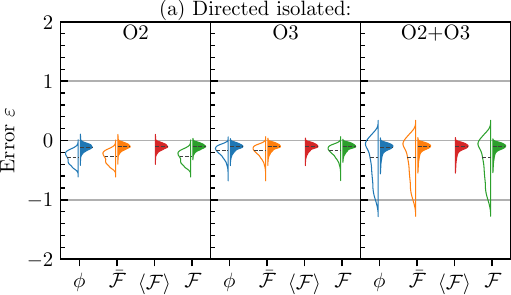}
  \includegraphics[width=\columnwidth]{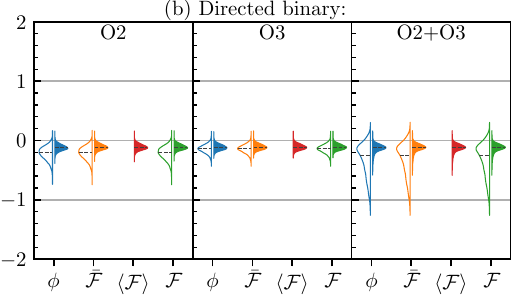}\\[\sskip]
  \includegraphics[width=\columnwidth]{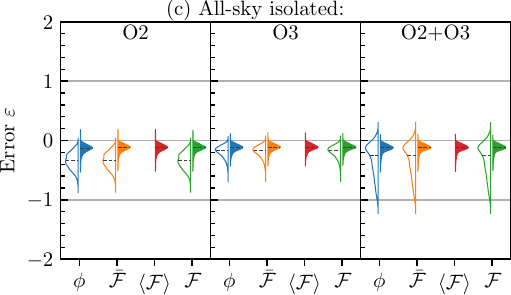}
  \includegraphics[width=\columnwidth]{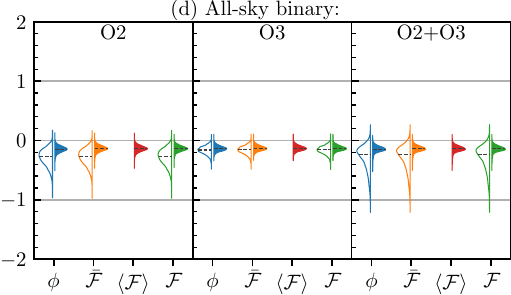}\\[\bskip]
  % --- Second 2x2 block: Weighted ---
  \textbf{II.\ Weighted $\sco{\F}_\weighted$-statistic ($\Tseg=\SI{10}{\day}$):}\\[\bskip]
  \includegraphics[width=\columnwidth]{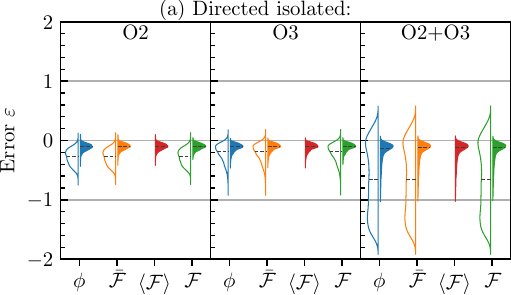}
  \includegraphics[width=\columnwidth]{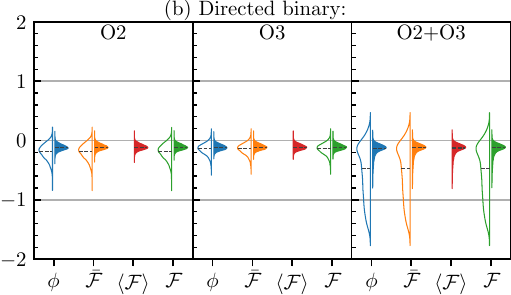}\\[\sskip]
  \includegraphics[width=\columnwidth]{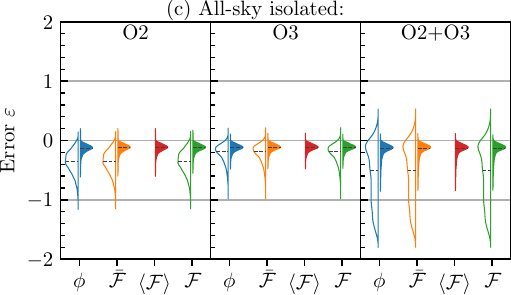}
  \includegraphics[width=\columnwidth]{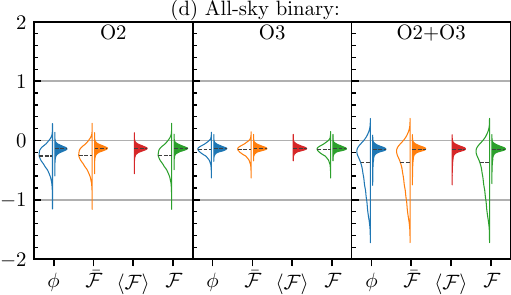}
  \caption{Same as Fig.~\ref{fig:coherent}, for semi-coherent searches with segments of duration
    $\Tseg = \SI{10}{\day}$.
    The top 2x2 block (I) is for the unweighted $\sco{\F}$-statistic, with
    phase metric $\sco{g}^\phi$, averaged $\F$-statistic metric $\sco{g}^{\meanF}$, marginalized $\F$-statistic metric
    $\sco{g}^{\margF}$, and the full $\F$-statistic metric $\sco{g}^{\F}$.
    The bottom 2x2 block (II) is for the weighted $\sco{\F}_\weighted$-statistic, with
    phase metric $\sco{g}_\weighted^\phi$, averaged $\F$-statistic metric $\sco{g}_\weighted^{\meanF}$, marginalized $\F$-statistic metric
    $\sco{g}_\weighted^{\margF}$, and the full $\F$-statistic metric $\sco{g}_\weighted^{\F}$.
  }
  \label{fig:semicoherent}
\end{figure*}
For the long-segment $\Tseg=\SI{10}{\day}$ semi-coherent searches (unweighted and weighted) the metric-prediction
error-distributions are shown in Fig.~\ref{fig:semicoherent}.
Similar to the coherent case, we see a general improvement of mismatch prediction accuracy by the new generalized
metrics, which is most pronounced for the combined O2+O3 dataset due to the different noise levels in these two runs,
which the idealized metric cannot take into account.

\subsubsection{Short-segment $\Tseg=\SI{0.1}{\day}$ semi-coherent search}

\begin{figure*}[tbp]
  % --- First 2x2 block: Unweighted ---
  \textbf{I.\ Unweighted $\sco{\F}$-statistic ($\Tseg=\SI{0.1}{\day}$):}\\[\bskip]
  \includegraphics[width=\columnwidth]{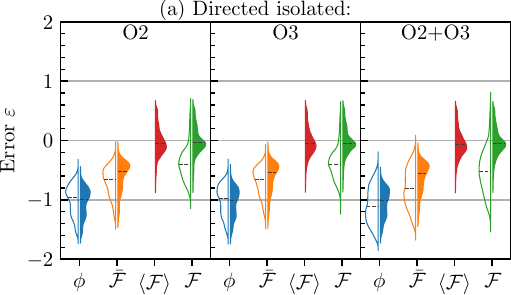}
  \includegraphics[width=\columnwidth]{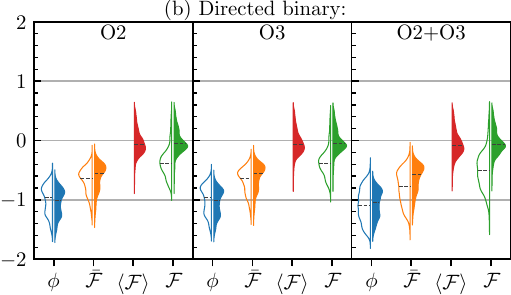}\\[\sskip]
  \includegraphics[width=\columnwidth]{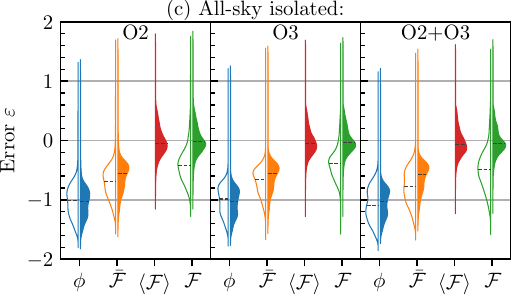}
  \includegraphics[width=\columnwidth]{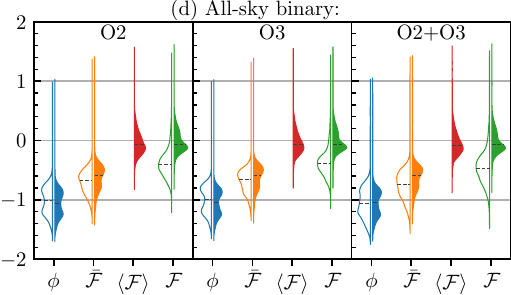}\\[\bskip]
  % --- Second 2x2 block: Weighted ---
  \textbf{II.\ Weighted $\sco{\F}_\weighted$-statistic ($\Tseg=\SI{0.1}{\day}$):}\\[\bskip]
  \includegraphics[width=\columnwidth]{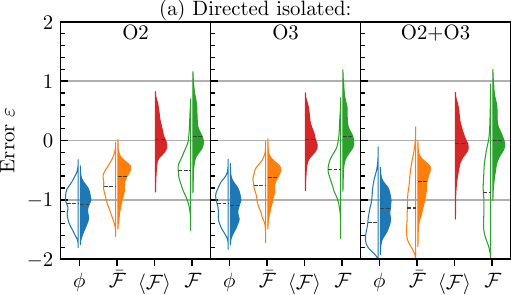}
  \includegraphics[width=\columnwidth]{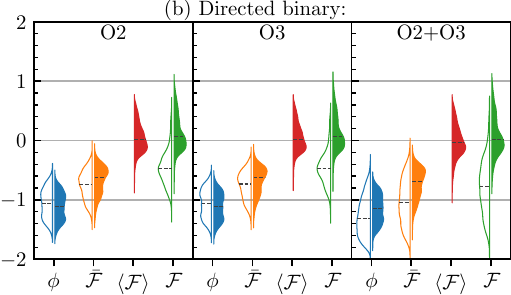}\\[\sskip]
  \includegraphics[width=\columnwidth]{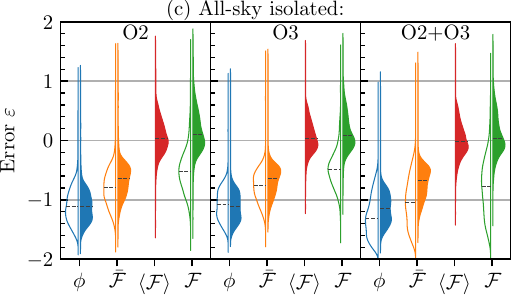}
  \includegraphics[width=\columnwidth]{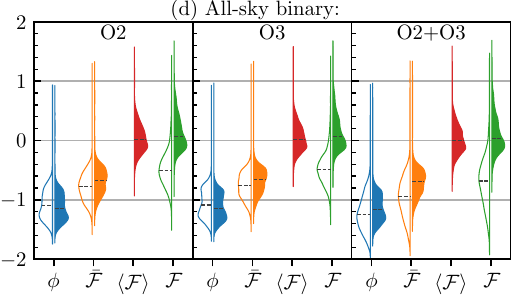}
  \caption{Same as Fig.~\ref{fig:coherent}, for semi-coherent searches with short segments of duration
    $\Tseg = \SI{0.1}{\day}$.
    The top 2x2 block (I) is for the unweighted $\sco{\F}$-statistic, with
    phase metric $\sco{g}^\phi$, averaged $\F$-statistic metric $\sco{g}^{\meanF}$, marginalized $\F$-statistic metric
    $\sco{g}^{\margF}$, and the full $\F$-statistic metric $\sco{g}^{\F}$.
    The bottom 2x2 block (II) is for the weighted $\sco{\F}_\weighted$-statistic, with
    phase metric $\sco{g}_\weighted^\phi$, averaged $\F$-statistic metric $\sco{g}_\weighted^{\meanF}$, marginalized $\F$-statistic metric
    $\sco{g}_\weighted^{\margF}$, and the full $\F$-statistic metric $\sco{g}_\weighted^{\F}$.
  }
  \label{fig:semicoherent2}
\end{figure*}
The error-distributions for the short-segment $\Tseg=\SI{0.1}{\day}$ semi-coherent searches are presented in
Fig.~\ref{fig:semicoherent2}, which display a number of interesting features:
(1) Much bigger differences between the different metric types, and especially a poor performance of the phase $g^\phi$
and averaged $g^{\meanF}$ metrics, which is likely related to effects discussed previously in \cite{Metric} (e.g., see Figs.~3
and 4).
We also note that the $\F$-statistic itself is somewhat sub-optimal for such short segments as first discussed
in \cite{PhysRevD.105.124007}.
(2) A marked improvement from the averaged $g^{\meanF}$ metric to the new marginalized $g^{\margF}$ metric derived in
Sec.~\ref{sec:new-marginalized-f-metric}, which in all cases seems to perform virtually identically to the full
$\F$-statistic metric $g^{\F}$.
This is likely related to the degeneracy in gravitational-wave polarization information for short segments and the
related problems discussed in \cite{PhysRevD.105.124007}, which marginalization over physical priors seems to elegantly
avoid, see also \cite{prix2024analyticweaksignalapproximationbayes}.
(3) A larger spread of relative errors (and tails) especially in the all-sky cases (c),(d), likely due to the metric not
being constant even for small offsets $\Delta\lambda$, as shown for example in Fig.~14 of \cite{Metric}.  This can
probably be avoided by choosing better sky coordinates, e.g., see \cite{wette_flat_2013}, but is beyond the scope of
this paper, and unrelated to the new metric approximation.

\subsection{Effects on metric ellipse volumes}
\label{sec:effect-metr-ellipse}

It is interesting to consider the effects of the generalized metric calculations of Sec.~\ref{sec:new} on the
metric-ellipse volume $V\propto 1/\sqrt{\det g}$, as this affects the total number of templates $\mathcal{N}$ required
at a certain mismatch via Eq.~\eqref{eq:nstar}.
We therefore compute the metric ellipse-volume ratio $r$ between the previous (idealized) $g\sub{prev}$ and the new
(generalized) $g\sub{new}$ metrics as
\begin{equation}
  r \equiv \frac{V\sub{prev}}{V\sub{new}} = \sqrt{\frac{ \det g\sub{new} }{ \det g\sub{prev} }}\,.
  \label{eq:12}
\end{equation}
A larger metric-ellipse volume corresponds to a lower template density, and therefore to a smaller number of templates,
as seen in Eq.~\eqref{eq:nstar}, therefore $r<1$ corresponds to an expected reduction in number of templates with the
new metric implementation, while $r>1$ would indicate an expected increase.

We compute this ratio for the phase metric $g^\phi$, the average $\F$-statistic metric $g^{\meanF}$, and the
$\F$-statistic metric $g^{\F}$), while the new marginalized metric $g^{\margF}$ has no previous implementation to
compare to.

We only consider the semi-coherent cases here, as full-dataset fully-coherent searches are only computationally feasible
when following up small numbers of interesting candidates in small parameter-space regions, where the question of the
number of templates is irrelevant.

\begin{figure*}[tbp]
  % --- First 2x2 block: Unweighted ---
  \textbf{I.\ Unweighted $\sco{\F}$-statistic ($\Tseg=\SI{10}{\day}$):}\\[0.4em]
  \includegraphics[width=\columnwidth]{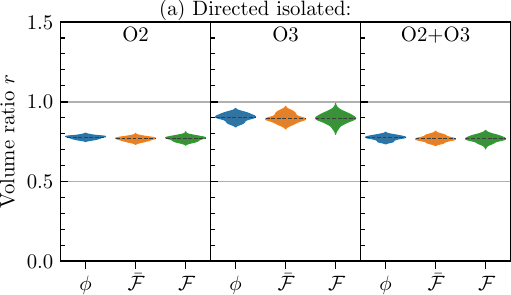}
  \includegraphics[width=\columnwidth]{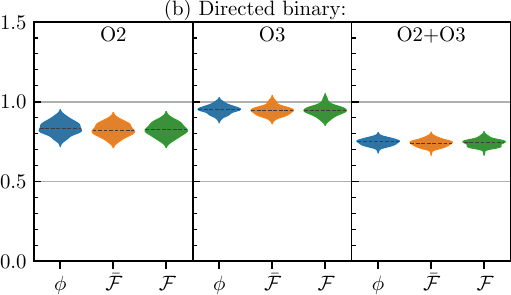}\\[0.2em]
  \includegraphics[width=\columnwidth]{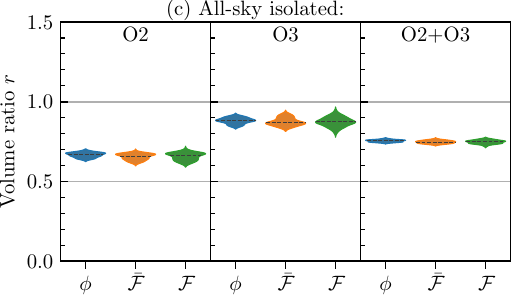}
  \includegraphics[width=\columnwidth]{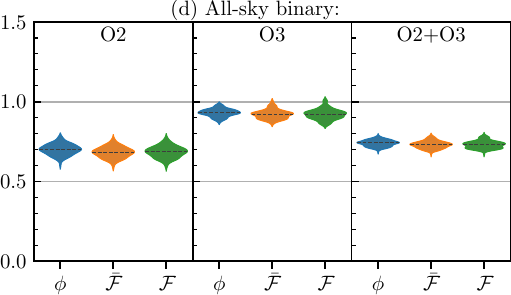}\\[0.4em]
  % --- Second 2x2 block: Weighted ---
  \textbf{II.\ Weighted $\sco{\F}_\weighted$-statistic ($\Tseg=\SI{10}{\day}$):}\\[0.4em]
  \includegraphics[width=\columnwidth]{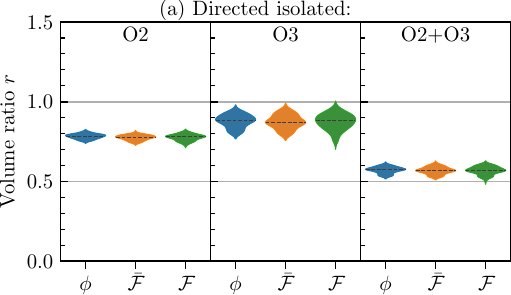}
  \includegraphics[width=\columnwidth]{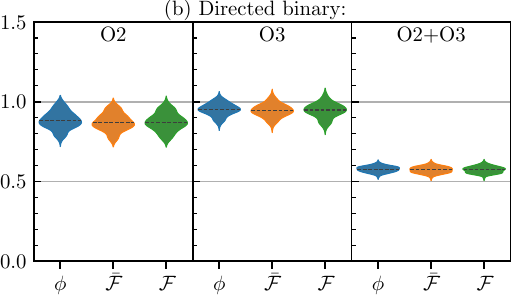}\\[0.2em]
  \includegraphics[width=\columnwidth]{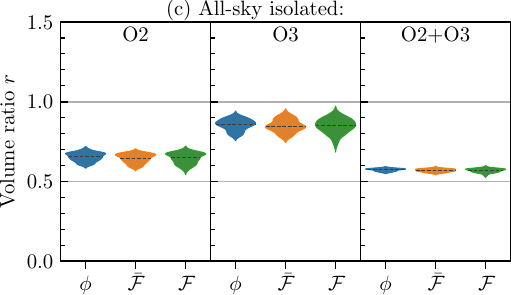}
  \includegraphics[width=\columnwidth]{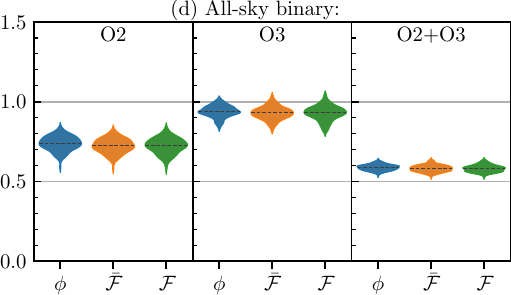}
  \caption{Distributions of metric ellipse-volume ratio $r$ of Eq.~\eqref{eq:12} for different
    semi-coherent $\F$-statistic searches with segments of duration $\Tseg = \SI{10}{\day}$:
    (a) directed isolated, (b) directed binary, (c) all-sky isolated, and (d) all-sky binary,
    using different data-sets: O2, O3 or O2+O3-combined, and metrics:
    phase metric $g^\phi$, averaged $\F$-statistic metric $g^{\meanF}$, and the full $\F$-statistic metric $g^{\F}$.
  }
  \label{fig:semicoherent3}
\end{figure*}
The results for the long-segment ($\Tseg=\SI{10}{\day}$) semi-coherent searches are shown in
Fig.~\ref{fig:semicoherent3}.  We observe very similar results for the different metric types, namely an overall
increase in metric ellipse volume (thereby reduction of expected number of templates) with the new generalized metrics.

\begin{figure*}[htbp]
  % --- First 2x2 block: Unweighted ---
  \textbf{I.\ Unweighted $\sco{\F}$-statistic ($\Tseg=\SI{0.1}{\day}$):}\\[0.4em]
  \includegraphics[width=\columnwidth]{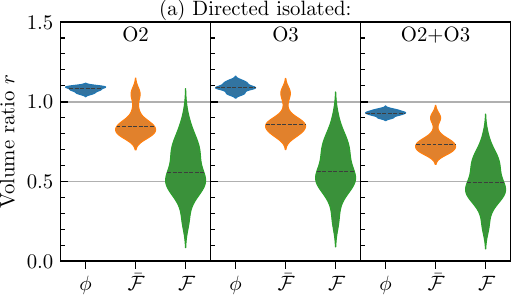}
  \includegraphics[width=\columnwidth]{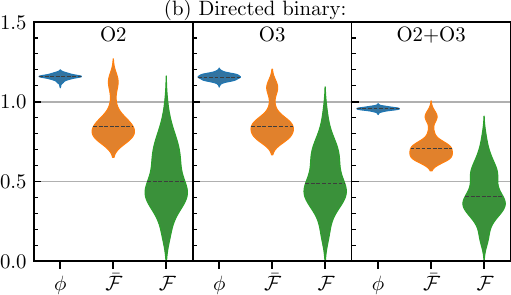}\\[0.2em]
  \includegraphics[width=\columnwidth]{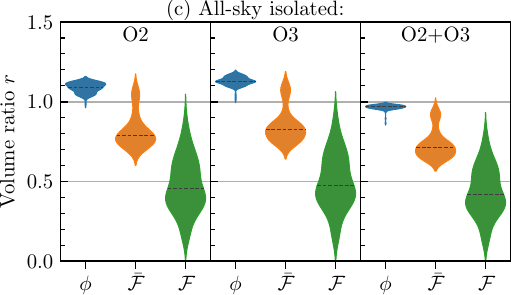}
  \includegraphics[width=\columnwidth]{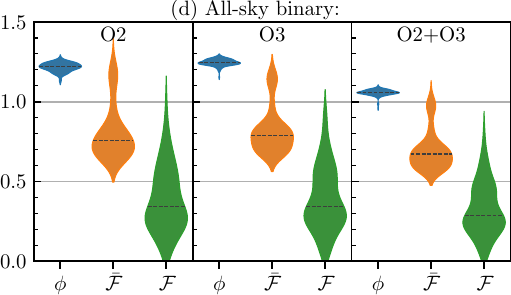}\\[0.4em]
  % --- Second 2x2 block: Weighted ---
  \textbf{II.\ Weighted $\sco{\F}_\weighted$-statistic ($\Tseg=\SI{0.1}{\day}$):}\\[0.4em]
  \includegraphics[width=\columnwidth]{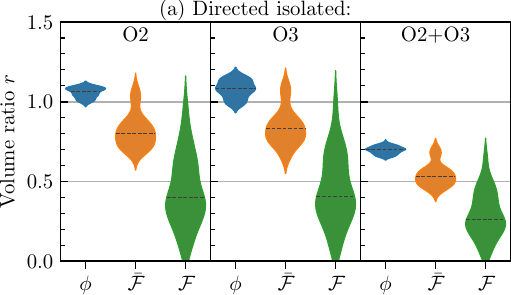}
  \includegraphics[width=\columnwidth]{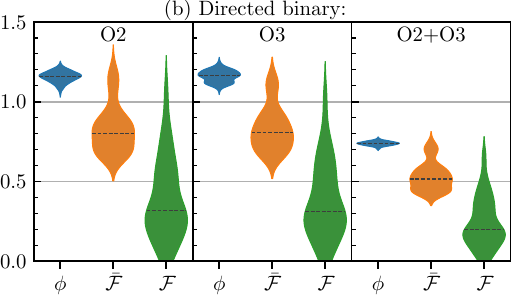}\\[0.2em]
  \includegraphics[width=\columnwidth]{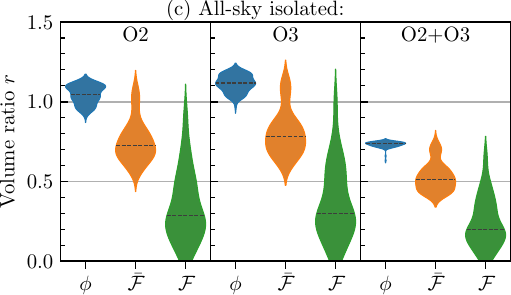}
  \includegraphics[width=\columnwidth]{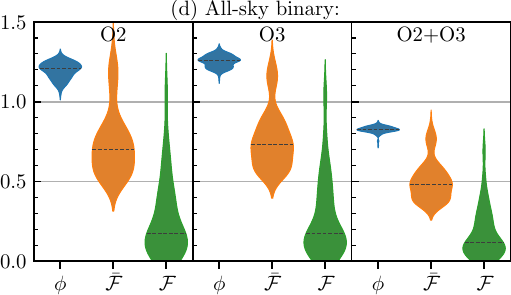}
  \caption{
    Distributions of metric ellipse-volume ratio $r$ of Eq.~\eqref{eq:12} for different
    semi-coherent $\F$-statistic searches with short segments of duration $\Tseg = \SI{0.1}{\day}$:
    (a) directed isolated, (b) directed binary, (c) all-sky isolated, and (d) all-sky binary,
    using different data-sets: O2, O3 or O2+O3-combined, and metrics:
    phase metric $g^\phi$, averaged $\F$-statistic metric $g^{\meanF}$, and the full $\F$-statistic metric $g^{\F}$.
    }
    \label{fig:semicoherent4}
  \end{figure*}
The short-segment semi-coherent searches are presented in Fig.~\ref{fig:semicoherent4}.
Here we observe large differences between the three different metric types, but as we saw in the corresponding
metric-prediction errors in Fig.~\ref{fig:semicoherent2}, the phase-metric $g^\phi$ accuracy is quite poor for these
short segments, and to some extent also the averaged $\F$-metric $g^{\meanF}$.
Therefore the result on the most accurate $\F$-statistic metric $g^{\F}$ is most significant here, which again shows a
(substantial) reduction in expected number of templates from the generalized metric calculations compared to the previous
idealized ones.

One can understand this general trend from the fact that taking into account of data-gaps and varying noise-floors
amounts to a reduction in the effective coherence time, which would decrease the metric resolution compared to the
idealized metrics that ignore these effects.

\section{Conclusions}
\label{sec:end}

In this paper we have generalized the expressions for the parameter-space metrics related to the $\F$-statistic.
We have shown that for realistic datasets with data gaps and a varying noise floor the idealized expressions can fail to
predict the shape and size of the $\F$-statistic mismatch correctly, and that the new generalized expressions are more
accurate.
We have presented a new metric approximation describing a polarization-marginalized $\F$-statistic metric $g^{\margF}$,
which seems to improve accuracy and robustness compared to the averaged $\F$-statistic metric $g^{\meanF}$.
We have further derived a more accurate semi-coherent metric that properly takes into account the variability of
signal power across segments by using weights, and we have also extended the metric expression for weighted
semi-coherent statistics.
We have shown that these new generalized metrics better predict the measured mismatches of the actual $\F$-statistic
when used on realistic datasets including gaps and varying noise-floors.

One downside of the new generalized metric expressions is a higher computational cost required to compute them, due to
the need to integrate SFT by SFT, which might complicate their application in cases where many different metrics need to
be computed, such as in an injection campaign with thousands of signals.

We have also found that the new metrics on realistic datasets (like the O2 and O3 observing runs) tend to reduce the
expected number of required templates and thereby the computing cost of searches using these metrics.

% Future work
Here we have only explored metrics for the classic $\F$-statistic, which was recently found to be quite sub-optimal
especially for short segments \cite{PhysRevD.105.124007,prix2024analyticweaksignalapproximationbayes}.
It would therefore be very interesting to investigate what metric expressions one obtains for these new short-segment
statistics, and how those would be related to the marginalized $\F$-statistic metric $g^{\margF}$.
It might also be interesting to investigate the parameter-space metric of the detection statistic that includes a
non-Gaussian line hypothesis such as \cite{2014PhRvD..89f4023K}.

\begin{acknowledgments}
This project has received funding from the European Union's Horizon 2020 research and innovation programme under the
Marie Sklodowska-Curie grant agreement number 101029058.

This work has utilized the ATLAS computing cluster at the MPI for Gravitational Physics Hannover.

This research has made use of data or software obtained from the Gravitational Wave Open Science Center
(\url{gwosc.org}), a service of the LIGO Scientific Collaboration, the Virgo Collaboration, and KAGRA \citep{GWOSC}.
\end{acknowledgments}

\appendix

\section{Weighted semi-coherent signal power}
\label{appendix:signalpower}

Let us look in more detail at the (perfect-match) weighted semi-coherent signal power $\sco{\rho}_{\weighted 0}^2$ of Eq.~\eqref{eq:rho2w},
by inserting the per-segment matched signal power $\rho_{0;\ell}^2$ of Eq.~\eqref{eq:rho2},
which leads to
\begin{align}
  \sco{\rho}_{\weighted 0}^2 &= \sum_\ell \weight_\ell \rho_{0;\ell}^2 \nonumber \\
                   &= h_0^2 \sum_\ell \weight_\ell \gamma_\ell \left( \alpha_1 A_\ell + \alpha_2 B_\ell + 2 \alpha_3 C_\ell \right).
                     \label{eq:38}
\end{align}
Using Eq.~\eqref{eq:18} we can see that semi-coherent data factor $\sco{\gamma}$
is self-consistently defined as
\begin{equation}
  \label{eq:19}
  \sco{\gamma} \equiv \S^{-1}\,\Tdata = \sum_\ell \gamma_\ell\,,
\end{equation}
where $\S$, defined in Eq.~\eqref{eq:54}, and $\Tdata$ now refer to the full dataset over all segments.
We can therefore write the weighted signal power Eq.~\eqref{eq:38} in the form
\begin{equation}
  \label{eq:20}
  \sco{\rho}_{\weighted 0}^2 = h_0^2 \,\sco{\gamma}\, \sco{g}_\weighted^2\,,
\end{equation}
where we defined the (weighted) semi-coherent geometric response function as
\begin{equation}
  \label{eq:21}
  \sco{g}_\weighted^2 \equiv \alpha_1 \sco{A}_\weighted + \alpha_2 \sco{B}_\weighted + 2\alpha_3\sco{C}_\weighted,
\end{equation}
in terms of the (weighted) semi-coherent antenna-pattern coefficients
\begin{align}
  \label{eq:AntennaW}
  \sco{\gamma} \sco{A}_\weighted  &\equiv \sum_\ell \weight_\ell \gamma_\ell A_\ell, \nonumber \\
  \sco{\gamma} \sco{B}_\weighted  &\equiv \sum_\ell \weight_\ell \gamma_\ell B_\ell,\\
  \sco{\gamma} \sco{C}_\weighted  &\equiv \sum_\ell \weight_\ell \gamma_\ell C_\ell\,.\nonumber
\end{align}
As expected, the above weighted expressions reduce to the standard semi-coherent forms when dropping the weights $\weight_\ell$.

\section{Amplitude-marginalizing the coherent $\F$-statistic metric}
\label{appendix:averaged}

In order to marginalize the $\F$-statistic metric $g^{\F}_{ij}$ of Eq.~\eqref{eq:metric-fstat} over $\cosi,\psi$,
we start by writing it more compactly as
\begin{equation}
  \label{eq:22}
  g^{\F}_{ij}(\cosi,\psi;\dop) = \sum_{k=1}^{3}m_{k;ij}(\dop)\frac{\al_k(\cosi,\psi) }{g^2(\cosi,\psi;\dop)}\,,
\end{equation}
where for convenience of notation we defined $\al_1\equiv\alpha_1$, $\al_2\equiv\alpha_2$ and $\al_3\equiv 2\alpha_3$.
With the marginalization integrals over $\cosi,\psi$ of Eq.~\eqref{eq:4}, we can write
\begin{equation}
  \label{eq:23}
  g^{\margF}_{ij}(\dop) \equiv \avg{g^{\F}_{ij}}_{\cosi,\psi} = \sum_{k=1}^3 m_{k;ij}(\dop) K_k(\dop)\,,
\end{equation}
in terms of the three marginalization integrals $K_k$, for which we propose the approximation
\begin{equation}
  \label{eq:24}
  K_k(\dop) \equiv \avg{\frac{\al_k}{g^2}}_{\cosi,\psi} \approx \frac{\avg{\al_k}_{\cosi,\psi}}{\avg{g^2}_{\cosi,\psi}}\,,
\end{equation}
replacing the average of the ratio by the ratio of averages.
We will test the accuracy of this approximation numerically in Sec.~\ref{sec:quant-marg-appr}.
We can use the known averages (e.g., \cite{prix:_cfsv2}) of
$\avg{\alpha_1}_{\cosi,\psi} = \avg{\alpha_2}_{\cosi,\psi} = 2/5$ and $\avg{\alpha_3}_{\cosi,\psi}=0$, to obtain
$\avg{g^2}_{\cosi,\psi}=\frac{2}{5} (A + B)$ from Eq.~\eqref{eq:g}, and therefore
\begin{equation}
  \label{eq:25}
  K_1 \approx K_2 \approx \frac{1}{A + B}\,,\quad K_3 \approx 0\,.
\end{equation}
Inserting this approximation into Eq.~\eqref{eq:23} we now find
\begin{equation}
  \label{eq:26}
  g^{\margF}_{ij} \approx \frac{m_{1;ij} + m_{2;ij}}{A+ B}\,,
\end{equation}
as stated in Eq.~\eqref{eq:metric-marginalized}.

\section{Amplitude marginalization of semi-coherent metrics}
\label{appendix:semicohw}

Here we consider the amplitude-marginalization of the most general semi-coherent metric $\sco{g}^{\F}_{\weighted;ij}$
of Eq.~\eqref{eq:13b} for (weighted) $\F$-statistics.
We can write this amplitude-marginalized metric as
\begin{align}
  \label{eq:28}
  \sco{g}^{\margF}_{\weighted;ij} &\equiv \avg{\sco{g}^{\F}_{\weighted;ij}}_{\cosi,\psi}\nonumber\\
                                  &= \frac{1}{\Nseg}\sum_\ell \avg{W_\ell\,g^{\F}_{ij,\ell}}_{\cosi,\psi}\nonumber\\
  &= \sum_{k=1}^3\sum_{\ell=1}^{\Nseg} \weight_\ell\, m_{k;ij,\ell} \avg{\frac{\rho_{0;\ell}^2}{\sco{\rho}_{\weighted 0}^2}\frac{\al_k}{g_\ell^2}}_{\cosi,\psi}\,,
\end{align}
where in the last line we substituted the metric weights $W_\ell$ of Eq.~\eqref{eq:11} and the per-segment $\F$-statistic metric
$g^{\F}_{ij,\ell}$ of Eq.~\eqref{eq:22}.
Using the explicit signal-power expression Eq.~\eqref{eq:rho2} with Eq.~\eqref{eq:20}, this leads to
\begin{align}
  \label{eq:29}
  \sco{g}^{\margF}_{\weighted;ij} &= \sum_{k\ell} \weight_\ell\,m_{k;ij,\ell}\,\frac{\gamma_\ell}{\sco{\gamma}} K_{\weighted;k}\,,
\end{align}
where we defined (similar to the coherent case given by Eq.~\eqref{eq:24}) the marginalization integrals $K_{\weighted;k}$, for which
we propose the approximation
\begin{equation}
  \label{eq:30}
  K_{\weighted;k} \equiv \avg{\frac{\al_k}{\sco{g}_\weighted^2}}_{\cosi,\psi} \approx
  \frac{\avg{\al_k}_{\cosi,\psi}}{\avg{\sco{g}_\weighted^2}_{\cosi,\psi}}\,,
\end{equation}
with the result in the same form as in Sec.~\ref{appendix:averaged}:
\begin{equation}
  \label{eq:31}
  K_{\weighted;1} \approx K_{\weighted;2} \approx \frac{1}{\sco{A}_\weighted + \sco{B}_\weighted}\,,\quad K_{\weighted;3} \approx 0\,.
\end{equation}
Putting this together with Eq.~\eqref{eq:29} we obtain the approximation
\begin{align}
  \label{eq:32}
  \sco{g}^{\margF}_{\weighted;ij} &\approx \sum_\ell \weight_\ell \frac{\gamma_\ell}{\sco{\gamma}}\,
                                    \frac{m_{1;ij,\ell} + m_{2;ij,\ell}}{\sco{A}_\weighted + \sco{B}_\weighted}\nonumber\\
                                  &\approx \sum_\ell \left(\weight_\ell \frac{\gamma_\ell}{\sco{\gamma}}
                                    \frac{A_\ell + B_\ell}{\sco{A}_\weighted + \sco{B}_\weighted}\right)\,
                                    g^{\margF}_{ij,\ell}\,,
\end{align}
where in the last line we used the coherent marginalized $\F$-statistic metric $g^{\margF}_{ij,\ell}$ of Eq.~\eqref{eq:metric-marginalized}.
We can see that the weight factor multiplying $g^{\margF}_{ij,\ell}$ corresponds to the (marginalized) metric segment weights
$\mweight_\ell$ given in Eq.~\eqref{eq:15}, namely
\begin{equation}
  \label{eq:33}
  \mweight_\ell \equiv k'\,w_\ell \gamma_\ell\,( A_\ell + B_\ell)\,,
\end{equation}
with normalization
\begin{equation}
  \label{eq:34}
  k' = \frac{\Nseg}{\sco{\gamma}\,(\sco{A}_\weighted + \sco{B}_\weighted)},
\end{equation}
satisfying the normalization condition $\sum_\ell \mweight_\ell = \Nseg$.
Therefore we can write
\begin{equation}
  \label{eq:35}
  \sco{g}^{\margF}_{\weighted;ij} \approx \frac{1}{\Nseg}\sum_\ell \mweight_\ell\,g^{\margF}_{ij,\ell}\,,
\end{equation}
as stated in Eq.~\eqref{eq:14}.

In order to see that this form of the result also holds true for the phase metric, we first note that
\begin{align}
  \label{eq:36}
  \avg{W_\ell}_{\cosi,\psi} &= \Nseg \weight_\ell\avg{\frac{\rho_{0,\ell}^2}{\sco{\rho}_{\weighted 0}^2}}_{\cosi,\psi}\nonumber\\
  % &= \Nseg \weight_\ell \frac{\gamma_\ell}{\sco{\gamma}}\avg{\frac{g_\ell^2}{g_\weighted^2}}\nonumber\\
                            &\approx \Nseg \weight_\ell \frac{\gamma_\ell}{\sco{\gamma}}\frac{\avg{g_\ell^2}}{\avg{g_\weighted^2}}\nonumber\\
                            &= \Nseg\frac{\weight_\ell\gamma_\ell\,(A_\ell+B_\ell)}{\sco{\gamma}\,(\sco{A}_\weighted + \sco{B}_\weighted)}\nonumber\\
  &= \mweight_\ell\,,
\end{align}
and therefore the amplitude-marginalized weighted phase-metric is also approximated in the form
\begin{align}
  \label{eq:37}
  \sco{g}^{\avg{\phi}}_{\weighted;ij} &\equiv \avg{g_{\weighted;ij}^\phi}_{\cosi,\psi} = \frac{1}{\Nseg}\sum_\ell \avg{W_\ell} g^\phi_{ij,\ell}\nonumber\\
  &\approx \frac{1}{\Nseg}\sum_\ell \mweight_\ell\,g^\phi_{ij,\ell}\,,
\end{align}
as stated in Eq.~\eqref{eq:14}.

\section{Testing the amplitude marginalization approximation}
\label{sec:quant-marg-appr}

We perform numerical tests for the two cases of the coherent approximation given by Eq.~\eqref{eq:25} of Sec.~\ref{appendix:averaged}
as well as the semi-coherent weighted case of Eq.~\eqref{eq:31} of Sec.~\ref{appendix:semicohw}. In these tests we compare the true values, computed numerically,
to the approximations.

Given $K_3$ is approximated as $0$, we cannot compute a standard relative error, instead we chose a common approximation ``scale'' of
$K_0 \equiv 1/(A+B)$ to compute the relative importance of the absolute deviations in the sum in Eq.~\eqref{eq:23}, namely
\begin{equation}
  \label{eq:27}
  \relerr_1 \equiv \frac{K_1 - K_0}{K_0},\quad
  \relerr_2 \equiv \frac{K_2 - K_0}{K_0},\quad
  \relerr_3 \equiv \frac{K_3}{K_0}\,.
\end{equation}
We have made the same definitions for the semi-coherent weighted case, substituting $K_k$ and $\relerr_k$ by $K_{\mathrm{w};k}$ and $\relerr_{\mathrm{w};k}$.

\begin{figure*}[htbp]
  \centering
  \includegraphics[width=\columnwidth]{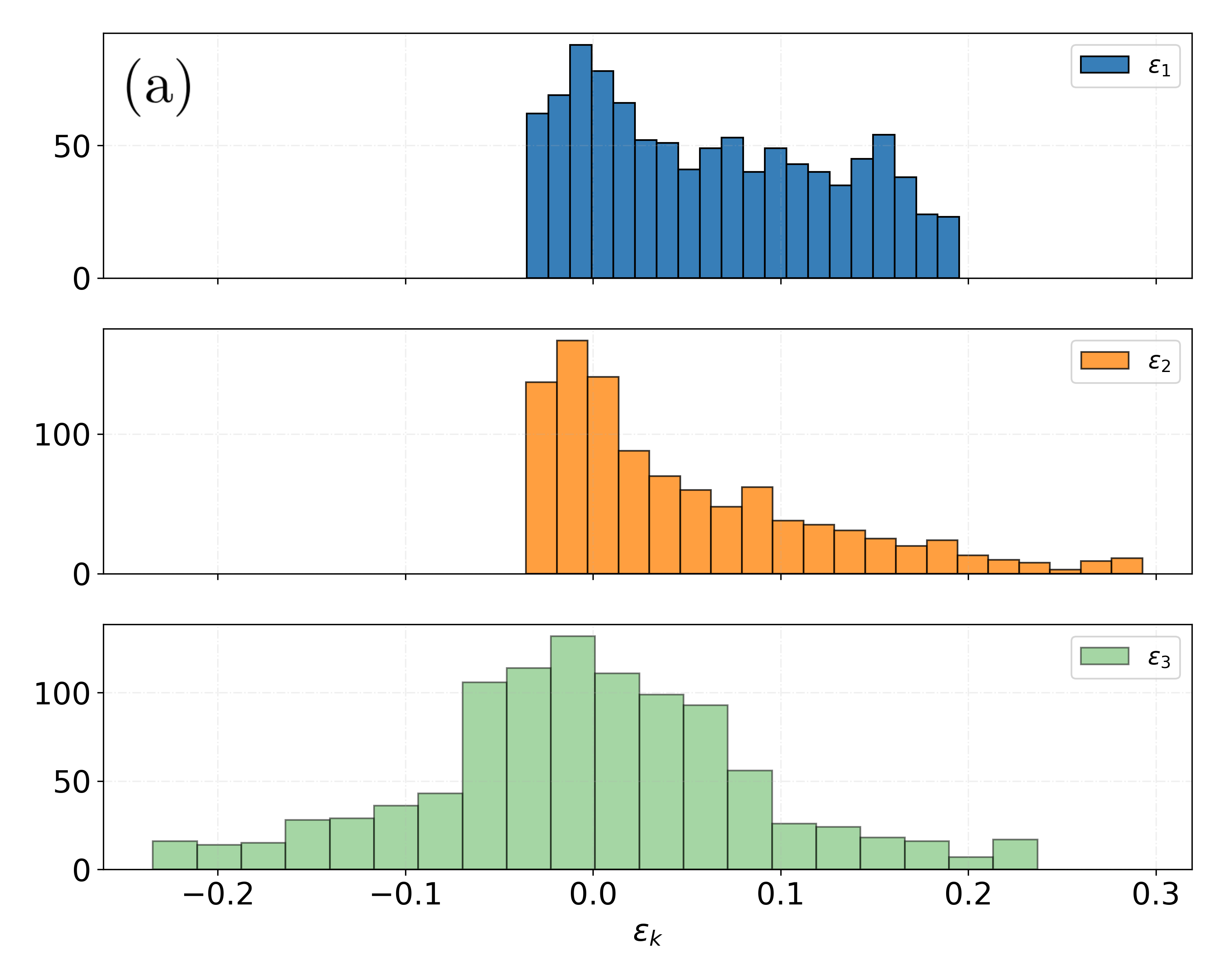}
  \includegraphics[width=\columnwidth]{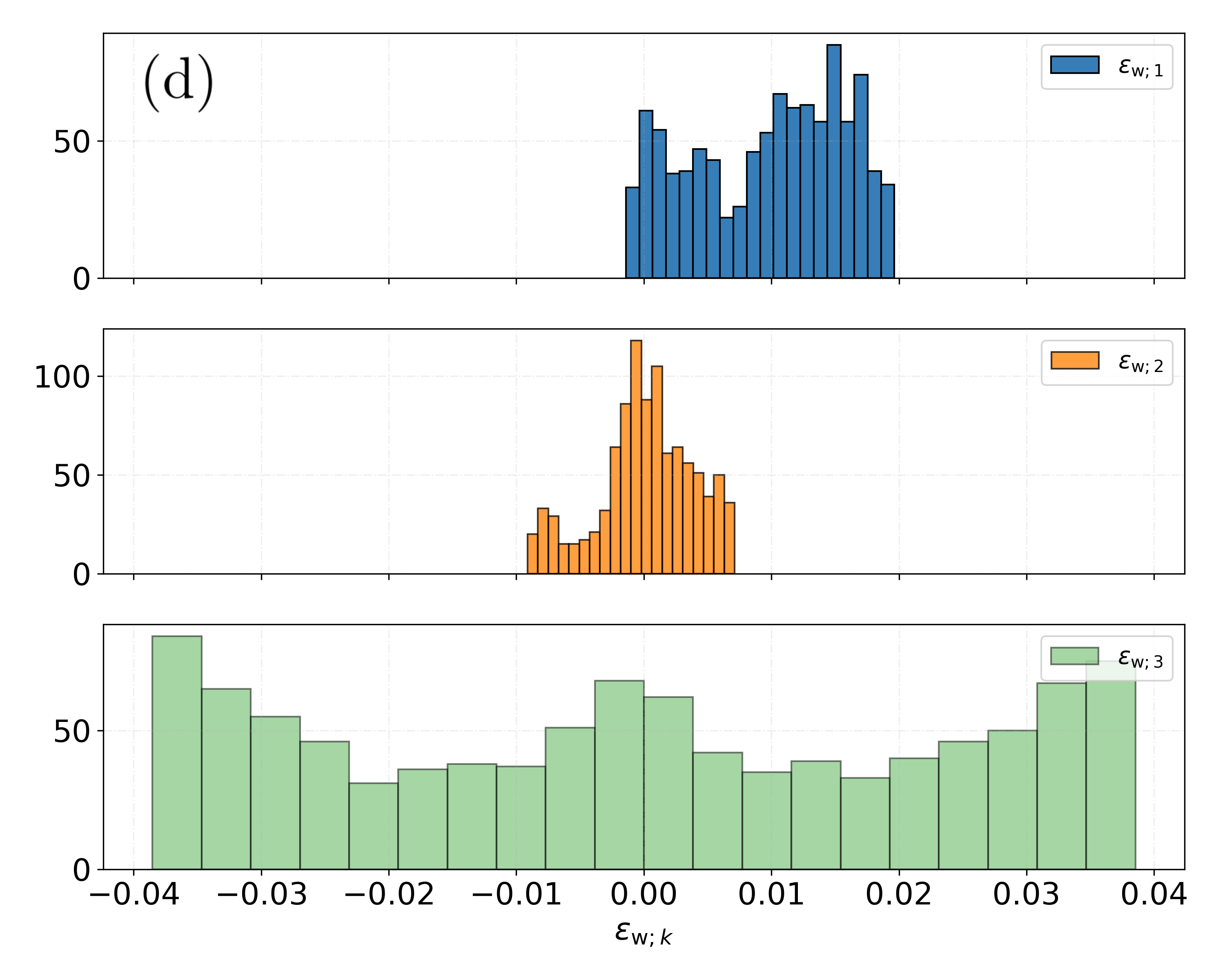}
  \includegraphics[width=\columnwidth]{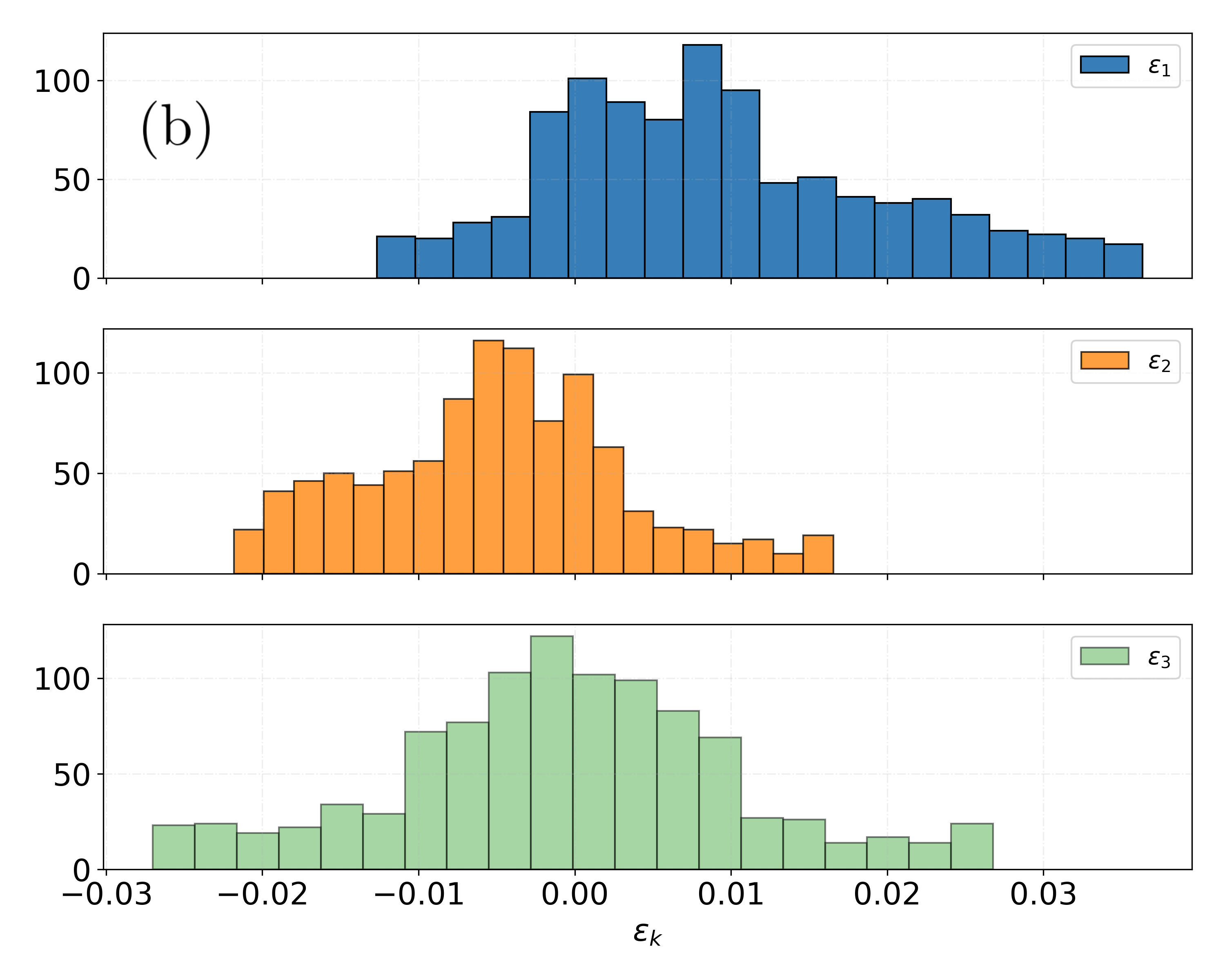}
  \includegraphics[width=\columnwidth]{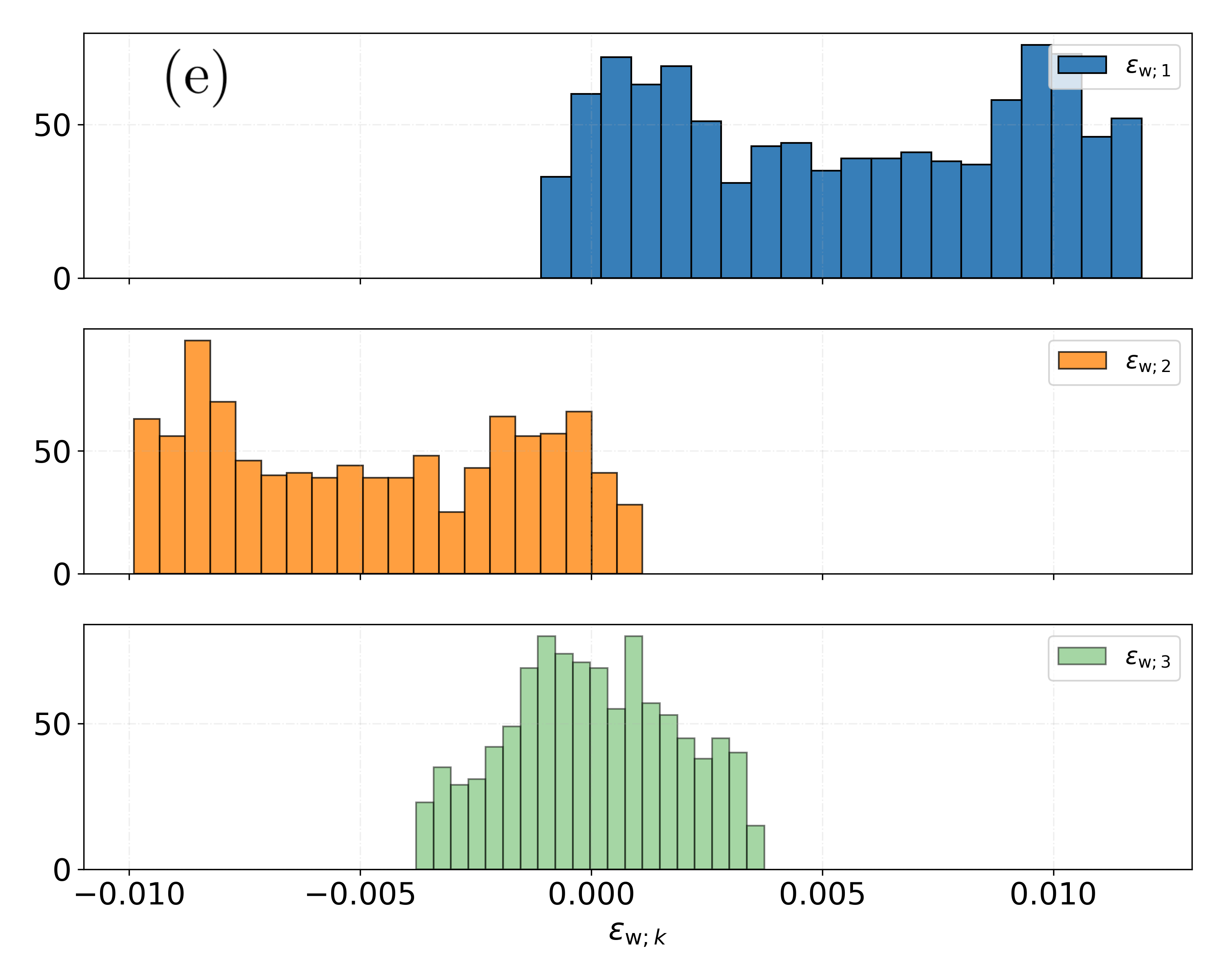}
  \includegraphics[width=\columnwidth]{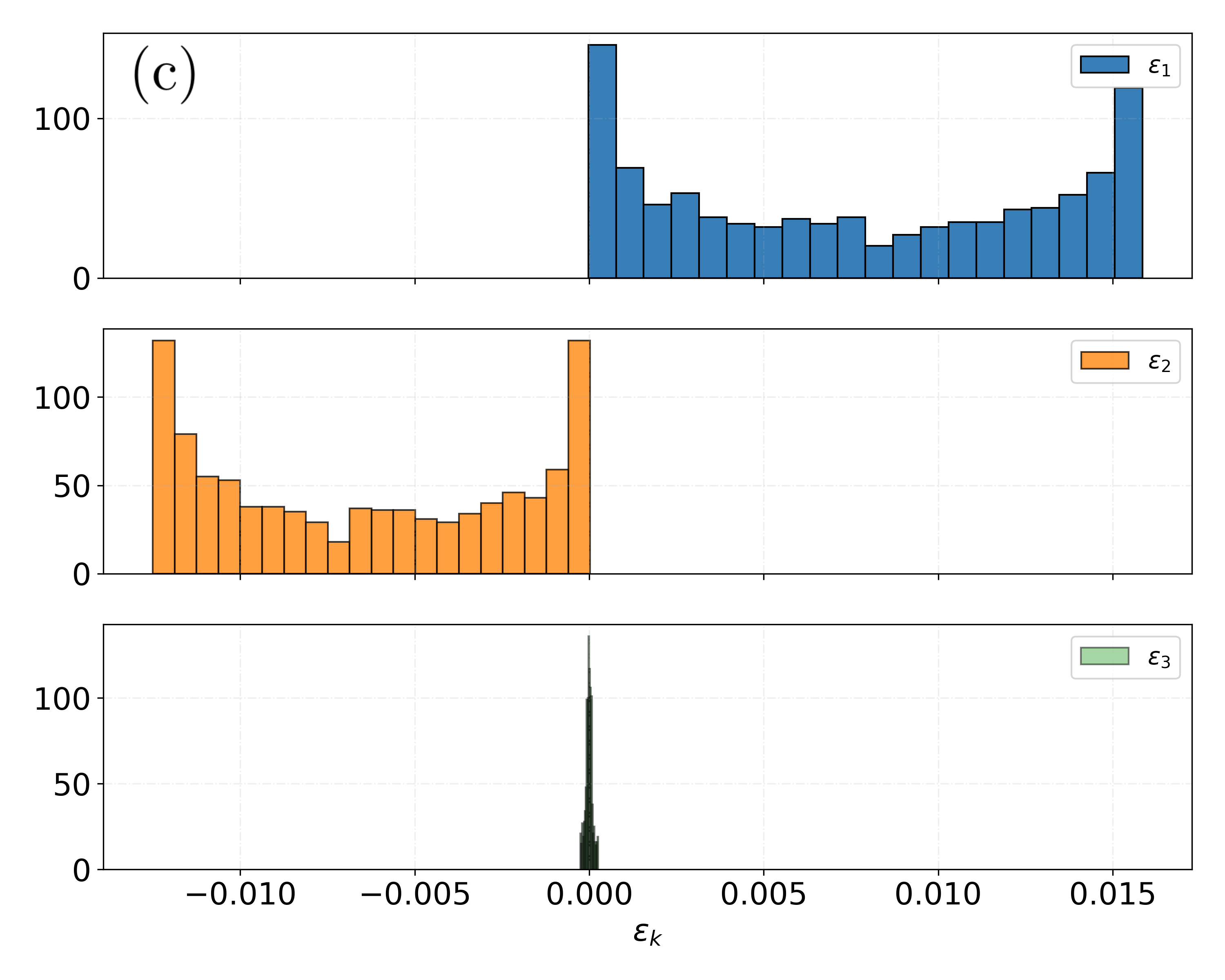}
  \includegraphics[width=\columnwidth]{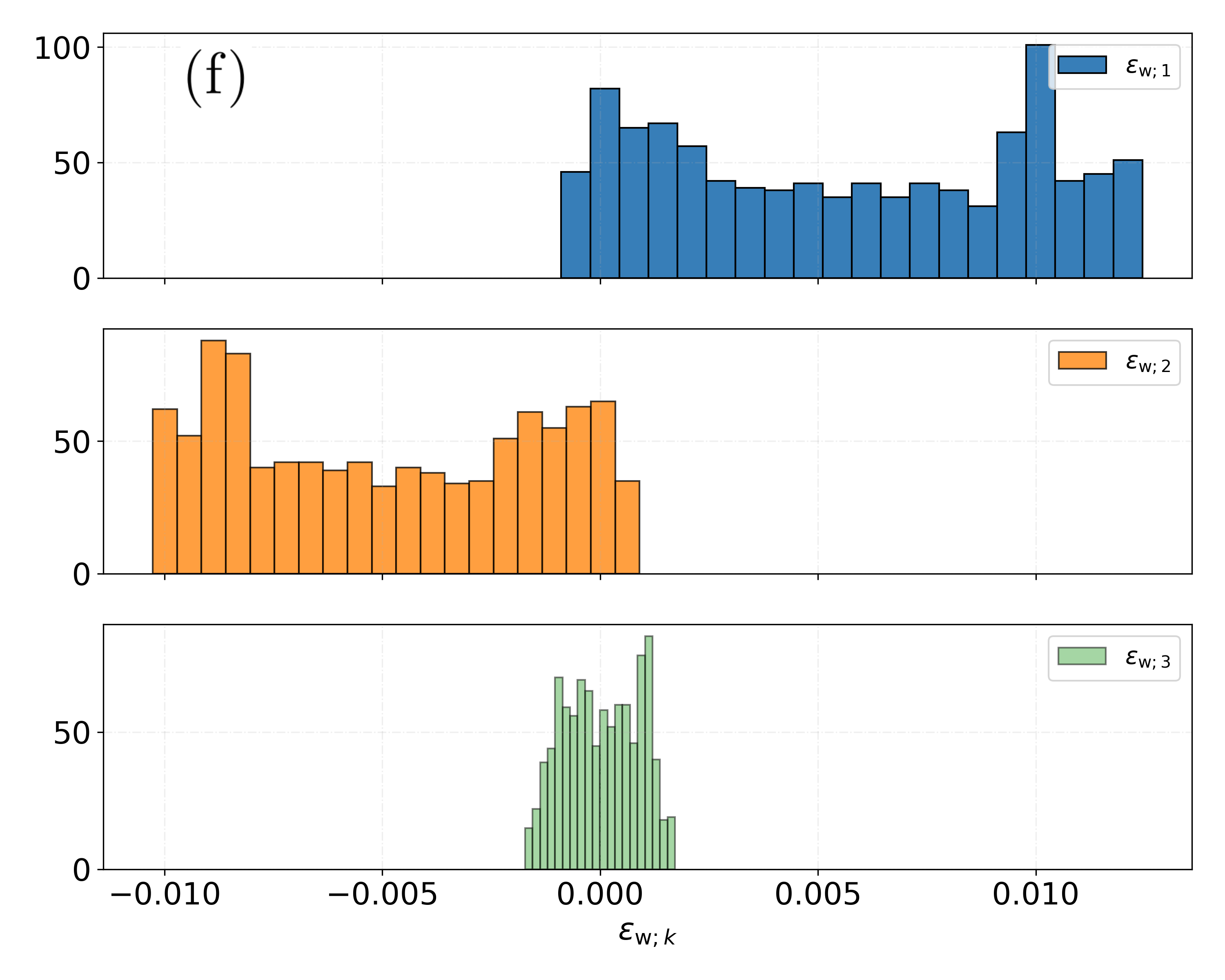}
  \caption{Histogram showing $\relerr_k$, defined in Eq.~\eqref{eq:27}. The \num{1000} simulated signals of each histogram are isotropically distributed over the sky. For the left column, the antenna pattern matrix coefficients have been obtained with \SI{1800}{s} SFTs without gaps from the H1 and L1 detectors with constant and equal noise floors covering (from top row to bottom) 0.25 days (a), 1.2 days (b), and 146 days (c). For the right column, the antenna pattern matrix coefficients have been obtained with \SI{1800}{s} SFTs from the H1 and L1 detectors belonging to the O3 observing run that have been modified with a time-domain cleaning algorithm \cite{PhysRevD.105.022005} with a segment time of (from top row to bottom) $T\seg = 900$ s (d), $T\seg = 1.2$ days (e), and $T\seg = 12.1$ days (f).}
  \label{fig:Ki}
\end{figure*}

Fig.~\ref{fig:Ki} shows histograms of these relative deviations computed over randomly chosen sky-positions for different data spans.
We see that this approximation tends to only introduce small errors on the order of a few percent for observation times of a day and longer.
However, for shorter observations such as $\Tdata=1/4$~days, we see that this approximation is less precise, and a different approach might be needed, which is postponed to future work.

\bibliography{Paper}

\end{document}

%% file: defs.tex
\newcommand{\F}{\mathcal{F}}
\newcommand{\A}{\mathcal{A}}
\newcommand{\M}{\mathcal{M}}
\renewcommand{\S}{\mathcal{S}}
\newcommand{\dop}{\lambda}
\newcommand{\ddop}{\Delta\!\dop}
\newcommand{\cosi}{\cos\iota}
\newcommand{\cosisq}{\cos^2\!\iota}
\newcommand{\sco}[1]{\widehat{#1}}
\newcommand{\expect}[1]{E\left[ #1 \right]}
\newcommand{\Ord}[1]{\mathcal{O}\left(#1\right)}
\newcommand{\vn}{\hat{n}}
\newcommand{\sub}[1]{_{\mathrm{#1}}}
\newcommand{\seg}{\sub{seg}}
\newcommand{\Tseg}{T\seg}
\newcommand{\Tdata}{T\sub{data}}
\newcommand{\Nseg}{N\seg}
\newcommand{\Nsft}{N\sub{SFTs}}
\newcommand{\Tsft}{T\sub{SFT}}
\newcommand{\avg}[1]{\left\langle #1\right\rangle}
\newcommand{\mean}[1]{\bar{#1}}
\newcommand{\meanF}{\mean{\F}}
\newcommand{\asini}{a\sub{p}}
\newcommand{\argp}{\omega}
\newcommand{\tasc}{t\sub{asc}}
\newcommand{\ecc}{e}
\newcommand{\Porb}{P\sub{orb}}
\newcommand{\sig}{\sub{s}}

\newcommand{\relerr}{\varepsilon}
\newcommand{\weight}{w}
\newcommand{\weighted}{{\mathrm{w}}}
\newcommand{\mweight}{\varpi}
\newcommand{\margF}{\avg{\F}}
\newcommand{\al}{\underline{\alpha}}